\documentclass[useAMS,usenatbib]{mn2e}

\usepackage{subfigure}
\usepackage{graphicx}
\usepackage{times}
\usepackage{tabularx}
\usepackage{natbib}
\usepackage{url} 
\usepackage{xcolor}
\usepackage{caption}
\usepackage{amsmath}
\usepackage{float}
\usepackage{multirow}
\usepackage{morefloats}

\newcommand{\enzo}{{\it {\small ENZO}}}

\newcommand{\dd}{\mathrm{d}}
\newcommand{\Mpc}{\mathrm{Mpc}}
\newcommand{\Msun}{\mathrm{M}_{\odot}}

\newcommand{\kpc}{\mathrm{kpc}}

\newcommand{\cm}{\mathrm{cm}}
\newcommand{\km}{\mathrm{km}}
\newcommand{\sek}{\mathrm{s}}

\newcommand{\G}{\mathrm{G}}
\newcommand{\K}{\mathrm{K}}

\newcommand{\para}{\mathrm{para}}
\newcommand{\erg}{\mathrm{erg}}
\newcommand{\Hz}{\mathrm{Hz}}
\newcommand{\MHz}{\mathrm{MHz}}
\newcommand{\GHz}{\mathrm{GHz}}

\newcommand{\nuobs}{\nu_{\mathrm{obs}}}
\newcommand{\Pburn}{P_{\mathrm{pol}}}

\newcommand{\Ppara}{P_{\mathrm{\perp}}}
\newcommand{\Pperp}{P_{\mathrm{\parallel}}}

\newcommand{\Cspec}{C_{\mathrm{spec}}}

\newcommand{\rad}{\mathrm{rad}}

\newcommand{\m}{\mathrm{m}}

\newcommand{\RM}{\mathrm{RM}}
\newcommand{\me}{m_\mathrm{e}}
\newcommand{\mpr}{m_\mathrm{p}}
\newcommand{\gram}{\mathrm{g}}

\begin{document}
 \definecolor{myred}{rgb}{1,0,0} 
 \definecolor{myblue}{rgb}{0,0,1}

 \title[Polarisation of Radio Relics]{Polarisation of Radio Relics in Galaxy Clusters}
 \author[D. Wittor, M. Hoeft, F. Vazza, M. Br\"{u}ggen]{D. Wittor$^{1,2,3}$\thanks{%
 E-mail: denis.wittor@unibo.it}, M. Hoeft$^{4}$, F. Vazza$^{1,2,3}$, M. Br\"{u}ggen$^{3}$, P. Dom\'inguez-Fern\'andez$^{3}$\\
 $^{1}$Dipartimento di Fisica e Astronomia, Universita di Bologna, Via Gobetti 93/2, 40122, Bologna, Italy \\
 $^{2}$ INAF, Istituto di Radioastronomia di Bologna, via Gobetti 101, I-41029 Bologna, Italy \\
 $^{3}$ Hamburger Sternwarte, Gojenbergsweg 112, 21029 Hamburg, Germany \\
 $^{4}$ Th\"uringer Landessternwarte, Sternwarte 5, 07778 Tautenburg, Germany }
 \date{Accepted ???. Received ???; in original form ???}
 \maketitle

 \begin{abstract}
   Radio emission in the form of giant radio relics is observed at the periphery of galaxy clusters. This non-thermal emission is an important tracer for cosmic-ray electrons and intracluster magnetic fields. One striking observational feature of these objects is their high degree of polarisation  which provides information on the magnetic fields at the relics' positions. In this contribution, we test if state-of-the-art high resolution cosmological simulations are able to reproduce the polarisation features of radio relics. Therefore, we present a new analysis of high-resolution cosmological simulations to study the polarisation properties of radio relics in detail. In order to compare our results with current and future radio observations, we create mock radio observations of the diffuse polarised emission from a massive galaxy clusters using six different projections, for different observing frequencies and for different telescopes. Our simulations suggest that, due to the effect of Faraday rotation, it is extremely difficult to relate the morphology of the polarised emission for observing frequencies below $1.4 \ \GHz$ to the real magnetic field structure in relics. We can reproduce the observed degree of polarisation and also several small-scale structures observed in real radio relics, but further work would be needed to reproduce some large-scale spectacular features as observed in real radio relics, such as the "Sausage" and the "Toothbrush" relics. 
 \end{abstract}
 \label{firstpage}
 \section{Introduction}\label{sec::intro}
 Radio observations detect diffuse radio emission in form of radio relics and radio halos in galaxy clusters \citep[e.g.][]{2008SSRv..134...93F,2019SSRv..215...16V}. Radio halos fill the clusters' central region and they are believed to be connected to the turbulence in the intracluster medium (ICM). Radio relics are found at the cluster periphery co-located with shock waves observed in X-rays \citep[e.g.][]{2005ApJ...627..733M,2016MNRAS.463.1534B}. Hence, it is assumed that shock (re)acceleration of cosmic-ray electrons produces radio relics \citep[e.g.][and references therein]{1998A&A...332..395E,2007MNRAS.375...77H,2014IJMPD..2330007B,2017MNRAS.470..240N}. It is widely accepted that diffusive shock acceleration (DSA) can accelerate the observed radio emitting electrons \citep[e.g.][]{1978ApJ...221L..29B}, yet the lack of detected $\gamma$-ray emission, a bi-product of proton acceleration \citep[e.g.][]{2014ApJ78718A}, questions the viability of the DSA model for typical cluster shocks \citep[e.g.][]{va14relics,va15relics}. This triggered the exploration of new scenarios in which the role of the magnetic field topology at shock fronts can control the efficiencies of electron and proton acceleration \citep[e.g.][and references therein]{2014ApJ...794...46C,Guo_eta_al_2014_II}, as we tested in detail in \citet{2017MNRAS.464.4448W}. More recently, \citet{2019arXiv190700966B} have found evidences that an efficiency $\geq 100 ~\%$  is required to explain the observed radio power of relics with low Mach number (i.e. $ \leq 2$).\\
\begin{figure*}
  \includegraphics[width = \textwidth]{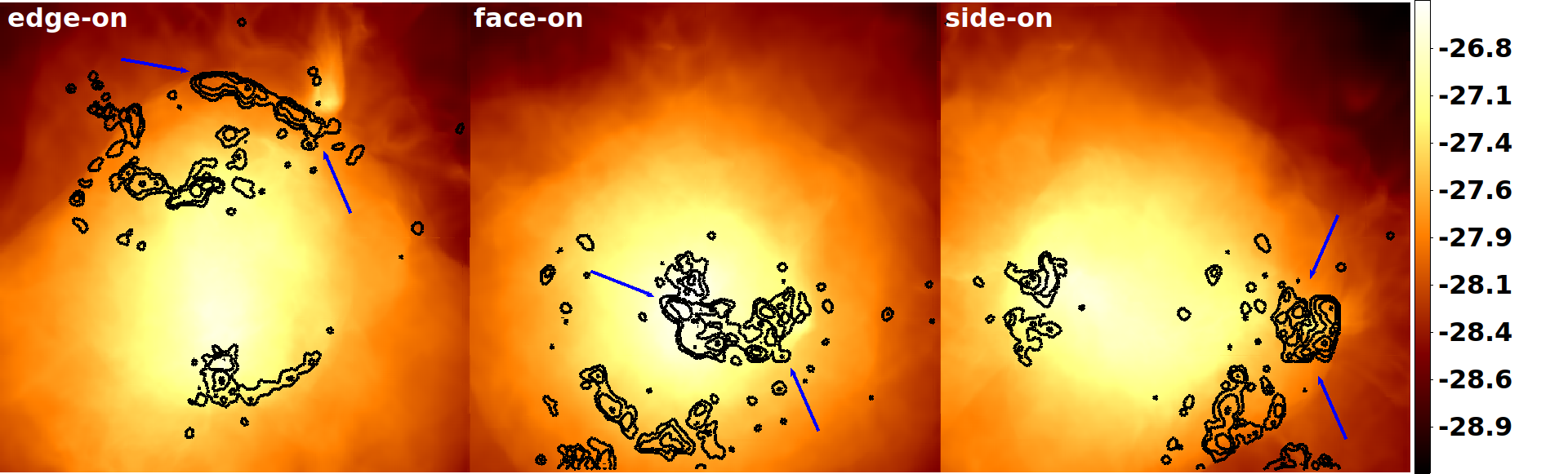} 
  \caption{Projected baryonic density (colour) overlayed with the radio contours at $\nuobs = 1.4 \ \GHz$. The density is given in units of $\log_{10} (\gram/\cm^3)$. The radio contours are plotted at $4 \cdot [10^{24}, 10^{25}, 10^{26} \ \& \ 10^{27}] \cdot \erg/\sek/\Hz$ per pixel. The different panels show the projections along the three different lines-of-sight. The solid blue arrows mark the position of the brightest radio relic as seen in the different projections. (A coloured version is available in the online article.)}
  \label{fig::rtr}
\end{figure*}
 One striking observational feature of radio relics is their high degree of polarisation, which is in contrast to the almost complete absence of polarised emission in radio halos \citep[e.g.][and references therein]{2013A&A...554A.102G}. The mechanism producing this high degree of polarisation is still debated: either it is caused by regular large-scale magnetic fields or by the compression of tangled small-scale magnetic fields \citep{1980MNRAS.193..439L}.  The polarisation of radio relics has been studied by a number of authors\footnote{Tab. \ref{tab::relics_obs} provides an overview of known relics that have been studied in polarisation. We devote Sec. \ref{ssec::relic_obs} to describe these in more detail.} \citep[e.g.][]{2009A&A...494..429B,2010Sci...330..347V,2012A&A...546A.124V,2012MNRAS.426.1204K,2015MNRAS.453.3483D}. These observations have shown that the average degree of polarisation ranges from a few up to $\sim 60 \ \%$. 
 In some radio relics, the $B$-vector of the polarised synchrotron emission are perpendicular to the shock normal. However, some radio relics show a more complex polarisation structure. If the high degree of polarisation is caused by a regular large-scale magnetic field, then the alignment does not depend on the shock morphology. On the other hand, if the compression of small-scale magnetic fields is responsible for the high degree of polarisation, the magnetic field should always be perpendicular to the shock normal. A simple model \citep[e.g.][]{1998A&A...332..395E} shows that a moderately strong shock may explain the degree of polarisation \citep[see Fig. 12 in][]{2017A&A...600A..18K}. In a more elaborated scenario, \citet{2012MNRAS.423.2781I} showed that magnetic amplification due to compression at shocks can explain the high polarisation in radio relics \citep[see also][]{1998A&A...332..395E}, even though this was based on analytical arguments only and neglected the effect of the viewing angle. With increasing resolution and sensitivity, radio telescopes (e.g. \textit{Square Kilometre Array} and \textit{Very Large Array}) will be able to study the polarisation of radio sources with unprecedented detail \citep[e.g.][]{2015aska.confE..92J}.\\
 To our knowledge, only \citet{2013ApJ...765...21S} have studied the polarisation of radio relics in cosmological magneto-hydrodynamical simulations at an observing frequency of $\nuobs = 1.4 \ \GHz$. In their model, they assign all the radio emission to the shock front using Eq. 31 of \citet{2007MNRAS.375...77H} and they estimate the corresponding polarised emission using the notation of \citet{2009ApJ...693....1O}. Yet, ageing of cosmic-ray electrons and corresponding downstream effects on the radio emission is neglected. Their results show a significant variation of the polarisation, both, across and along the relic with a peak polarisation of approximately $75 \ \%$. Furthermore, they showed that the polarisation direction is less coherent if the relic is seen face-on instead of edge-on.\\
 In this contribution, we use \enzo \ cosmological magneto-hydrodynamical simulations of a merging galaxy cluster to analyse the radio properties of one radio relic seen in different projections. We compute the downstream emission of this relic at different frequencies and model the  polarised emission using the formalism derived by \citet{1966MNRAS.133...67B}. This work is meant to be a pilot study to explore the degree of realism of magnetic fields in high-resolution cosmological simulations and to present our test suite. \\
 This work is structured as follows: we complete this section by giving an overview of the available polarisation observations of radio relics. In  Sec. \ref{ssec::enzo}, we describe our simulation and analysis tools. In Sec. \ref{ssec::synchrotron} and \ref{ssec::polarisation}, we describe the models we use to compute the polarised emission. The results are presented in Sec. \ref{sec::results}. Sec. \ref{ssec::sample}, \ref{ssec::magneticfields} and \ref{ssec::rm}, give a detailed description of our simulated radio relic, its local magnetic field and its rotation measures, respectively. The results on the polarised emission are presented in Sec. \ref{ssec::properties}. We present our mock observations of the radio relic in Sec. \ref{ssec::telecsope}. In Sec. \ref{ssec::comparison}, where we highlight morphological features of our radio relic, as found in highly resolved observations. In Sec. \ref{ssec::magnetic}, we discuss how well the polarised emission represents the magnetic field structure at the relic. We summarise and conclude our work in Sec. \ref{sec::conclusion}.
\begin{table*}  
  \begin{tabular}{l||c|c|c|c|c|c|c|c|c}
  relic 	   & $z$ & Mach & $\nu \ [\MHz]$ & beam $[('')^2]$ & pol. frac.	& max pol. frac. & LLS $[\Mpc]$ & $d_{c}$ $[\Mpc]$ & reference \\ \hline \hline
1RXS J0603.3+4214  & 0.225  & 3.3-4.6 & 4900 & $7 \cdot 4.7$     & 0.13 & 0.60 & 1.9 &  -    & A \\ %
                   & 0.225  & -       & 8350 & $90 \cdot 90$     & 0.22 & 0.45 & 1.9 &  1.3  & B \\ %
	               & 0.225  & -       & 4850 & $159 \cdot 159$   & 0.15 & -    & 1.9 &  1.3  & B \\ %
Abell 746          & 0.232  & -   & 1382 & $23 \cdot 18$ & -  & 0.50 & 1.1  & 1.7  &  C \\ %
Abell 1240 north    & $0.195^a$  & 3.3  & 1425 & $18 \cdot 17$	  & 0.26 & 0.70 & 0.65 & 0.7  &  D \\ 
                    & 0.195  & $5.1^b$  & 3000 & $18.5 \cdot 14.5$ & 0.29 & 0.58 & -    & -    &  E \\
Abell 1240 south   & $0.195^a$ & 2.8  & 1425 & $18 \cdot 17$     & 0.29 & 0.70 & 1.25 & 1.1  &  D \\ 
                   & 0.195  & $4.0^b$  & 3000 & $18.5 \cdot 14.5$ & 0.16 & 0.40 & -    & -   &  E \\ 
  Abell 1612 & 0.179 & 2.47 & 8350 & $90 \cdot 90$   & 0.13 & 0.20 & 0.78 &  1.3  &  B \\ 
             & 0.179 & 2.47 & 4850 & $159 \cdot 159$ & 0.05 & -    & 0.78 &  1.3  &  B \\
Abell 2256 	       & 0.0594 &  -   & 1369 & $15 \cdot 14$ 		    & 0.20 			& 0.45 		& - & - &  F  \\ 
Abell 2345-1       & 0.177  & 2.8 & 1425 & $23 \cdot 16$ & 0.14 & 0.60 & 1.15 & 1.0  &  D \\ 
Abell 2345-2       & 0.177  & 2.2 & 1425 & $23 \cdot 16$ & 0.22 & 0.50 & 1.5  & 0.89 &  D \\ 
Abell 2744 	       & 0.308  & 2.05 & $3000^c$ & $10 \cdot 10$	  & 0.27 & 0.52 & 1.5  & 1.3  &  G \\ 
Abell 3376 east    & 0.046  & 3.31 & 1400 & $38 \cdot 26$		    & -		  	    & 0.30 	    & 0.9   & 1.0 &  H \\ 
Abell 3376 west    & 0.046  & 2.23 & 1400 & $37 \cdot 25$		    & -  			& 0.20		& 0.5   & 1.0 &  H \\ 
Abell 3411 	       & 0.1687  &  -  & 1400 & $48 \cdot 33$           &    - 			& 0.25 		& 1.9   & 1.3 &  I \\ 
Abell 3744 	       & 0.0381 &   -   & 1400 & $45 \cdot 45$		    & 0.33 		 	& -		    & 1.4 & 1.8 &  J \\ 
Abell 548b-A       & 0.04   &   -   & 1400 & $15 \cdot 30$		    & 0.30 			&  -		& 0.26 & 0.5 &  K \\
Abell 548b-B       & 0.04   &   -   & 1400 &	$15 \cdot 30$	        & 0.30 			&  -		& 0.31 & 0.43 &  K \\ 
Bullet             & 0.296  & 2.0-5.4 & 1400 & $10 \cdot 10^d$ 	& 0.01  & -     & 0.93 & 1.0 & L \\ 
                   & 0.296  & 2.0-5.4 & 1700 & $10 \cdot 10^d$  & 0.03 	& -		& 0.93 & 1.0 & L \\ 
                   & 0.296  & 2.0-5.4 & 2700 & $10 \cdot 10^d$ 	& 0.12 	& -		& 0.93 & 1.0 & L \\ 
CIZA J2242.8+5301  & 0.189  & 4.6  & 4900 & $5.2 \cdot 5.1$		& $0.55^e$ 			& 0.60 		& 2.0 & 1.5 & M \\ 
                   & 0.192  &  -    & 8350 & $90 \cdot 90$		& 0.29			& 0.55 		& 2.0 & 1.5 & B \\ 
                   & 0.192  &  -    & 4850 & $159 \cdot 159$	& 0.36 			& 0.45 		& 2.0 & 1.5 & B \\ 
El Gordo west 	   & 0.87  & 2.5 & 2100 & $38.4 \cdot 24.7$       & 0.33 		    & 0.67 		& 0.56 & -  & N \\ 
El Gordo east 	   & 0.87  &  -  & 2100 & $38.4 \cdot 24.7$       & 0.33		    & - 		& 0.27 & -  & N \\ 
MACS J0717+3745    & 0.55   & $2.7^f$ & 1365 & $5 \cdot 4$  & 0.08 & -    & 0.83 & 0.45 & O \\ 
                   & 0.55   &        -            & 4885 & $5 \cdot 4$  & 0.17 & 0.20 & -    & -    &  O \\ 
MACSJ1149.5+2223-W & 0.554  & 4.6  & 1450 & $23 \cdot 16$	        & 0.05       		& - 		& 0.76 & 1.14 & P \\ 
MACSJ1149.5+2223-E & 0.554  & 3.0  & 1450 & $23 \cdot 16$	        & 0.05        		& -		& 0.82 & 1.39 & P \\ 
MACSJ1752.0+4440-NE & 0.366 & 4.6  & 1666 & $23.5 \cdot 15.9$       & 0.20            & 0.40 		& 1.13 & 1.3 & P \\ 
MACSJ1752.0+4440-SW & 0.366 & 2.8  & 1666 & $23.5 \cdot 15.9$       & 0.10            & 0.40 		& 0.91 & 0.8 & P\\ 
PSZ1 G096.89+24.17 N& 0.3 & -  & 1400 & $15 \cdot 14$  & $0.15^g$  &  0.20   & 0.88  & 0.77 & Q \\ 
PSZ1 G096.89+24.17 S& 0.3 & -  & 1400 & $15 \cdot 14$  & $0.15^g$  &  0.20   & 1.419 & 1.145 & Q \\ 
PSZ1 G108.18-11.53 N & 0.335  & 2.2  & 1380 & $17 \cdot 13$	  & $0.20^h$ & 0.30 & 1.5  & 1.75 & R \\ 
PSZ1 G108.18-11.53 S & 0.335  & 2.33  & 1380 & $17 \cdot 13$  	  & $0.20^h$ & 0.30 & 1.3  & 1.28 & R \\ 
ZwCl 0008.8+5215 east & 0.103 & 2.2  & 1400 & $23.5 \cdot 17$  & -    & 0.25 & 1.4 &  -   & S\\ 
                      & 0.103 & 2.35 & 8350 & $90 \cdot 90$    & 0.22 & 0.26 & 1.4 & 0.9  & B\\ 
                      & 0.103 & 2.35 & 4850 & $159 \cdot 159$  & 0.13 & 0.22 & 1.4 & 0.9  & B\\ 
	                  & 0.104 & -   & $3000^c$ & $12 \cdot 14$ & 0.30 & 0.40 & - & - & T\\ 
ZwCl 0008.8+5215 west & 0.103 & 2.4  & 1400 & $23.5 \cdot 17$ & - & 0.10 & 0.29 & - & S\\ 
	                  & 0.104 & -   & $3000^c$  	& $12 \cdot 14$ & 0.18 & - & - & - & T\\ 
  \end{tabular}
  \caption{The table shows various properties of observed relics that have been studied in polarisation. The columns provide: the relic name, the redshift, the estimated Mach number, observing frequency and beam size. Followed by the average and the maximum  polarisation fraction of the relic. We also give the largest-linear-size (LLS) and the distance to the cluster centre ($d_c$) as found in literature. The table is complemented by the reference. We did not find, both, average and maximum degree of polarisation for all relics. The maximum values of the polarisation are difficult to measure so they should be taken with caution. The references given in the last column refer to the measurements of the polarisation fraction. All other quantities, if not stated otherwise, have been derived or quoted in these references. We put as an additional caveat that the Mach numbers might not have been computed in the same way. So, one should compare these values with caution as there is a bias in the Mach number estimations \protect{see Sec. \ref{ssec::sample}}. \\ Footnotes: a: taken from \protect{\citet{2018arXiv180610619G}}; b: we cite the Mach number derived from the integrated spectral index, while \citet{2018MNRAS.478.2218H} also use the injection spectral index to compute the Mach number. Using the injection spectral index, they obtain Mach numbers of $\sim 2.4$ and $\sim 2.3$ for Abell 1240-1 and Abell 1240-2 respectively; c: Both \protect{\citet{2017ApJ...838..110G}} and \protect{\citet{2017ApJ...845...81P}} only provide a frequency range, $2000-4000 \ \GHz$, hence, we assume the mean of $3000 \ \GHz$; d: we assume this value, following the caption of Fig. 3 in \protect{\citet{2015MNRAS.449.1486S}}; e: we give the arithmetic mean of the range, $0.5-0.6$, given in \protect{\citet{2010Sci...330..347V}}; f: taken from \protect{\citet{2017ApJ...835..197V}}; g: we give the arithmetic mean of the range, $0.1-0.2$, given in \protect{\citet{2014MNRAS.444.3130D}}; h: we give the arithmetic mean of the range, $0.1-0.3$, given in \protect{\citet{2015MNRAS.453.3483D}}. \\ References:  A: \protect{\citet{2012A&A...546A.124V}}; B: \protect{\citet{2017A&A...600A..18K}}; C: \protect{\citet{2011A&A...533A..35V}}; D: \protect{\citet{2009A&A...494..429B}}; E: \protect{\citet{2018MNRAS.478.2218H}}; F: \protect{\citet{2006AJ....131.2900C}}; G: \protect{\citet{2017ApJ...845...81P}}; H: \protect{\citet{2012MNRAS.426.1204K}}; I: \protect{\citet{2013ApJ...769..101V}}; J: \protect{\citet{2009AJ....137..145R}}; K: \protect{\citet{2006MNRAS.368..544F}}; L: \protect{\citet{2015MNRAS.449.1486S}}; M: \protect{\citet{2010Sci...330..347V}}; N: \protect{\citet{2014ApJ...786...49L}}; O: \protect{\citet{2009A&A...503..707B}}; P: \protect{\citet{2012MNRAS.426...40B}}; Q: \protect{\citet{2014MNRAS.444.3130D}}; R: \protect{\citet{2015MNRAS.453.3483D}}; S: \protect{\citet{2011A&A...528A..38V}}; T: \protect{\citet{2017ApJ...838..110G}}. }
  \label{tab::relics_obs}
\end{table*}
 \subsection{Polarisation Properties of observed Radio Relics}\label{ssec::relic_obs}
 While the statistical properties of the continuum emission from radio relics are relatively well-known \citep[e.g.][]{2017MNRAS.470..240N,2017arXiv171101347G}, far less is known about the properties of the polarised emission. Currently, there are only 20 galaxy clusters, hosting radio relics (half of which are double radio relics \citep[e.g.][]{2009A&A...494..429B,2014ApJ...786...49L}), that have been observed in polarisation. In Tab. \ref{tab::relics_obs}, we summarise the main properties of these observations. These relics have been detected at redshifts ranging from $z \approx 0.04$ to $z \approx 0.55$ and their largest-linear sizes (LLS) lie in the range of a few hundred $\kpc$ to a few $\Mpc$. Most of these relics are found at distances of $0.5 -  2 \ \Mpc$ from the cluster centre. We have not found Mach number estimates for all relics, but the available Mach numbers are all in the range of $M \approx 1.7 - 4.6$. \\
 The average degree of polarisation varies between a few percent up to $\sim 60 \ \%$, as in the case of CIZA J2242.8+5301 \citep[see][]{2010Sci...330..347V}. Locally, the observed degree of polarisation can be as high as $\sim 70 \ \%$ as in Abell 1240\footnote{However, the uncertainty of these measurements are fairly large, as Abell 1240 is very faint compared to other relics. \citet{2009A&A...494..429B} measured fluxes of $\sim 6 \ \mathrm{mJy}$ and $\sim 10 \ \mathrm{mJy}$ for the two relics, while \citet{2017MNRAS.472.3605L} measured a flux of $\sim 161 \ \mathrm{mJy}$ for the relic in CIZA J2242.8+5301.} \citep[see][]{2009A&A...494..429B}.\\
 The relic population shows a vast range of the orientation of the polarisation vectors. In the cluster CIZA J2242.8+5301, the polarisation $E$-vectors align perfectly with the shock normal across the entire length of the relic, $\sim 2 \ \Mpc$, \citep[see][]{2010Sci...330..347V} and other relics, such as Abell 2744 \citep[see][]{2017ApJ...845...81P} or ZwCl 0008.8+5215 \citep[see][]{2011A&A...528A..38V,2017ApJ...838..110G} show a similar behaviour. For a sufficiently strong shock, the magnetic field compression at the shock front could explain the alignment of the shock normal and the polarisation $E$-vector. On the other hand, such alignment is more difficult to achieve for a large-scale magnetic field, as will be shown in this work. \\
 The morphology of the polarised emission in other relics is more complex. The orientation of the polarisation vectors of radio relics, such as MACS J0717+3745 \citep[see][]{2009A&A...503..707B}, changes rapidly over the length of the relic. Therefore, their polarised structures are not uniform and they appear rather patchy. At the same time, there is no obvious alignment between the polarisation vector and the shock normal, making it unclear if the shock is simply too weak to align the magnetic field with the shock normal. \\
In other cases, such as Abell 3744, PSZ1 G096.89+24.17 or PSZ1 G108.18-11.53 \citep[see][respectively]{2012MNRAS.426.1204K,2014MNRAS.444.3130D,2015MNRAS.453.3483D}, the polarised emission is not detected across the whole relic and the morphology seems to be patchy as well. Thus, it is uncertain if the polarisation is uniform across the relic. \\
Finally, it is difficult to determine the shock direction in relics like Abell 548b or Abell 2256 \citep[see][respectively]{2006MNRAS.368..544F,2006AJ....131.2900C}, and hence one cannot correlate the polarisation vectors with the shock direction. The shock morphology might directly affect the polarisation morphology. A spherical-cap-type shock would produce aligned polarisation vectors, while a distorted shock structure could cause patchy polarised emission. \\
 In summary, the polarised emission of radio relics shows a wide range of properties. It seems that the observed degree of polarisation does not solely depend on one parameter, such as the telescope configuration (i.e. observing frequency and beam size) or the physical properties of the relic (i.e. distance to the cluster centre, Mach number, LLS or redshift). Clearly, the degree of polarisation increases with frequency (smaller wavelengths) as Faraday rotation, $\propto \RM  \lambda^2$, becomes non-relevant. This was shown by \citet{2015MNRAS.449.1486S} who observed the \textit{Bullet} cluster at different frequencies but keeping a constant beam size. On the other hand, beam depolarisation can reduce the degree of polarisation even at high frequencies. Several relics, such as CIZA J2242.8+5301, show this effect. \citet{2010Sci...330..347V} observed this relic at $4.9 \ \GHz$  obtaining an average polarisation fraction of $0.55$, while \citet{2017A&A...600A..18K} obtained a lower degree of polarisation at $4.85 \ \GHz$ and $8.35 \ \GHz$ but using larger beam sizes.\\%
\begin{table*}
  \begin{tabular}{l||c|c|c|c|c|c|c}
    projection & $S \ [10^3 \cdot \kpc^2]$ & $P_{1.4 \ \GHz} \ [10^{30} \ \erg/\sek/\Hz]$ & $\langle M \rangle$ & $ M_{(0.15-1.4) \ \GHz}$ & $\langle B [\mu \G] \rangle$ & $\langle B [\mu \G] \rangle_{1.4 \ \GHz}$ & LLS $[\Mpc]$ \\ \hline \hline
    edge-on  & 226 & 1.29 & 2.45 & 3.54 & 0.20 & 0.56 & 1.2 \\ \hline
    face-on  & 421 & 1.32 & 2.33 & 3.44 & 0.40 & 0.64 & 1.3\\ \hline
    side-on  & 248 & 1.30 & 2.37 & 3.48 & 0.28 & 0.35 & 0.7
  \end{tabular}
  \caption{The table shows the properties of our simulated radio relic. From left to right: the projection, the projected relic surface area, the radio luminosity at $1.4 \ \GHz$, the average Mach number (for Mach numbers $> 2$), the Mach number computed using the spectral index between $0.15 \ \GHz$ and $1.4 \ \GHz$, the average and radio-weighted average magnetic field strength at the shock front, and the largest linear size. The radio weighting was done at $1.4 \ \GHz$.}
  \label{tab::relics}
\end{table*}
\begin{figure*}
  \includegraphics[width = 0.95\textwidth]{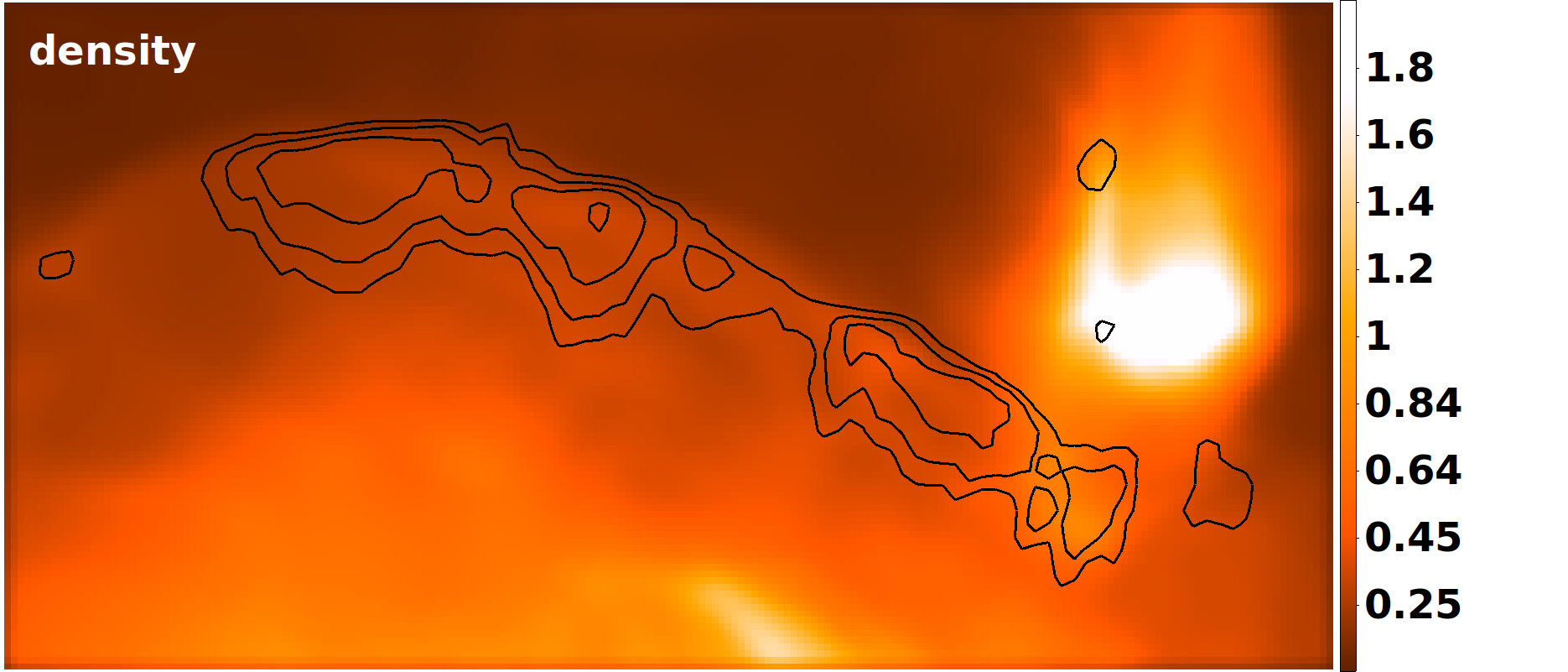} \\
  \includegraphics[width = 0.95\textwidth]{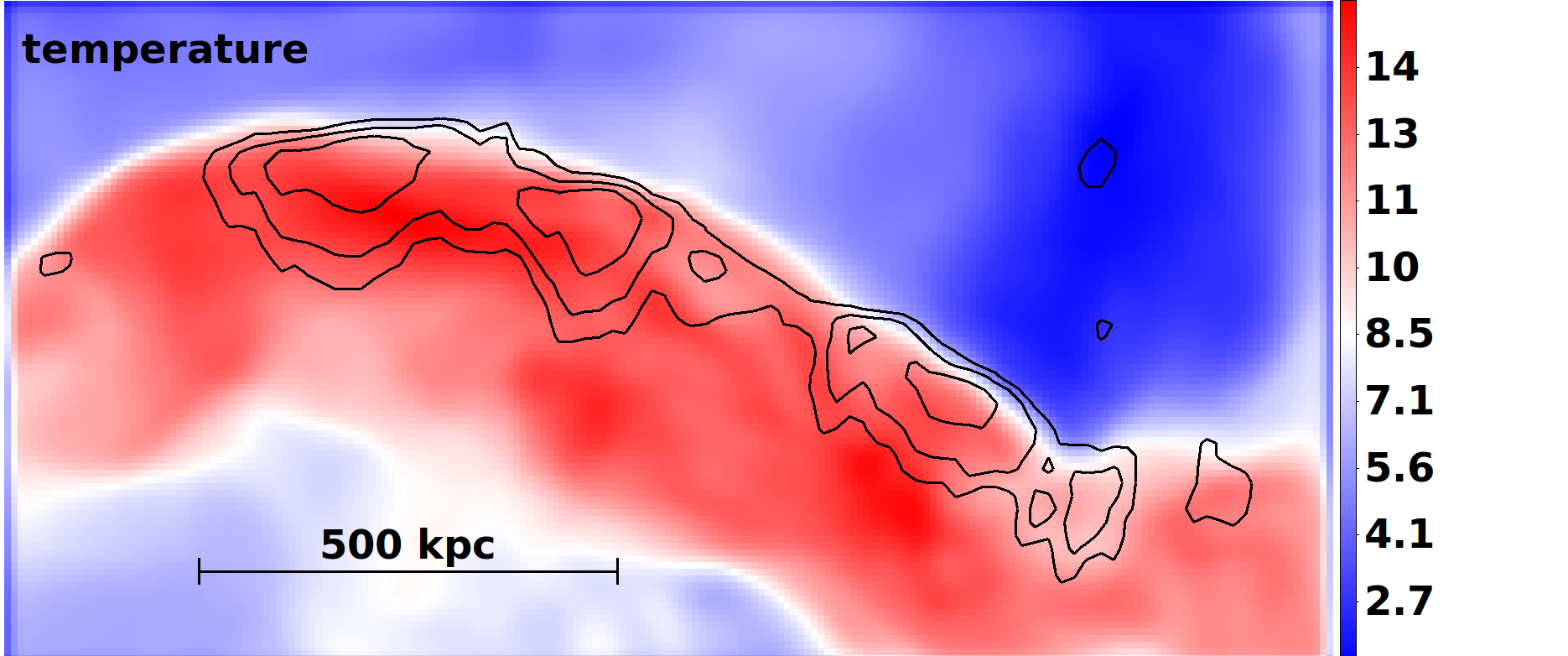} \\
  \includegraphics[width = 0.95\textwidth]{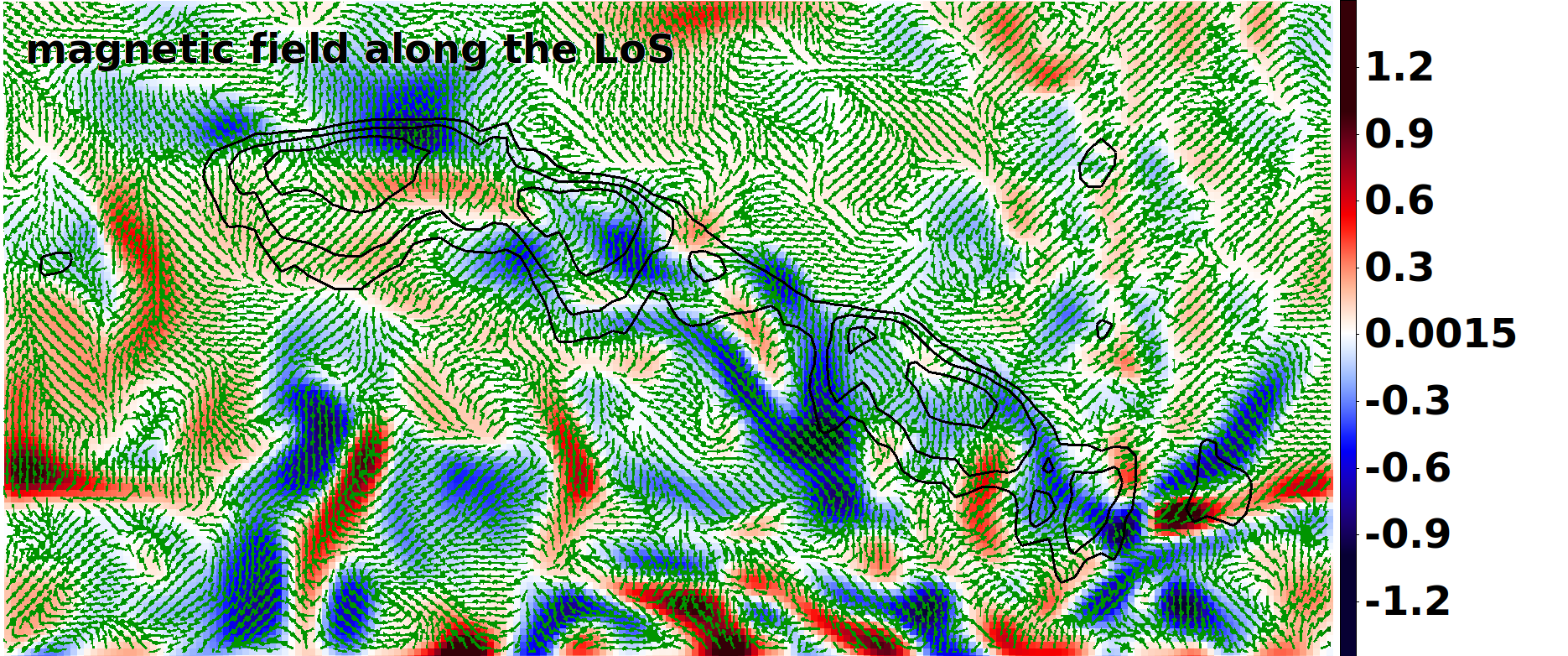} 
  \caption{Close-up view of the relic region. The black contours show the radio power at $1.4 \ \GHz$ at $4\cdot[10^{24}, 10^{25}, 10^{26} \ \& \ 10^{27}] \cdot \erg/\sek/\Hz/$. Top: projected density in $10^{-27} \ \gram/\cm^3$, middle: projected temperature in $10^{7} \ \K$ and bottom: slice through the magnetic field component along the line-of-sight in $\mu \G$ (colour) overlayed with the magnetic field vectors (green vectors) in the plane. The vectors are scaled to unity. (A coloured version is available in the online article.)}
  \label{fig::rel_environment}
\end{figure*}
 \section{Simulation setup}\label{sec::setup}
 \subsection{ENZO}\label{ssec::enzo}
 In this work, we run cosmological simulations with the \enzo-code \citep{ENZO_2014}, which uses an N-body particle-mesh solver to simulate the Dark Matter component \citep{1988csup.book.....H} and an adaptive mesh method to follow the baryonic component \citep{1989JCoPh..82...64B}. To solve for the magneto-hydrodynamical equations \citep[see Sec. 2.1 in][]{ENZO_2014}, we use the piecewise linear method \citep{1985JCoPh..59..264C} in combination with the hyperbolic Dedner cleaning \citep{2002JCoPh.175..645D}. \\
 We focus on analysing one massive, $\sim  10^{15} \ \Msun$ (at $z=0$), galaxy cluster drawn from a large sample of simulations, as detailed in  \citet[][]{2018MNRAS.474.1672V} and \citet{2019arXiv190311052D}.  When analysed with our shock finder (see below) this cluster hosts a few shock waves in its periphery that could produce prominent radio relics, with typical large-scale morphologies of real radio relics (see Fig. \ref{fig::rtr}). \\
 Our simulation starts from a root grid with a comoving size of $  \sim (260 \ \Mpc)^3$ that is sampled with $256^3$ cells and $256^3$ Dark Matter particles. We further refined a comoving volume of approximately $(25 \ \Mpc)^3$ centred around the galaxy cluster $2^8$ times, using 8 levels of AMR, for a final resolution of $\Delta x = 3.95 \ \kpc$. For the analysis we used the $7.9 \ \kpc$-reconstruction of the grid, as the relic region is lying at the border of the highest AMR region. \\
 The simulation ran from redshift $z=30$ to redshift $z = 0$. At redshift $z = 30$, we seeded a uniform primordial field, with a comoving value of $B_0 = 0.1\ \mathrm{n}\G$. For the various cosmological parameters we chose: $H_0 = 72.0 \ \km \ \sek^{-1} \ \Mpc^{-1}$, $\Omega_{\mathrm{M}} = 0.258$, $\Omega_{\mathrm{b}} = 0.0441$,  $\Omega_{\Lambda} = 0.742$ and $\sigma_8 = 0.8$ \\
 In order to find shock waves that are able to produce radio relics, we apply a velocity jump method following  \citet{2009MNRAS.395.1333V}. This approach measures the Mach numbers along the three coordinate axes of the simulation based on the 3-dimensional velocity information, and the final Mach number is computed as  $M = \sqrt{M_x^2 + M_y^2 + M_z^2}$. For further analysis, we stored the three components of the Mach number, as they provide the information of the propagation direction of the shock and of the shock normal in each shocked cell $\mathbf{n}_{\mathrm{shock}} = (M_x, M_y ,M_z)^{\mathsf{T}}$.
 \subsection{Synchrotron Emission}\label{ssec::synchrotron}
\begin{figure}
     \centering
     \includegraphics[width = 0.49\textwidth]{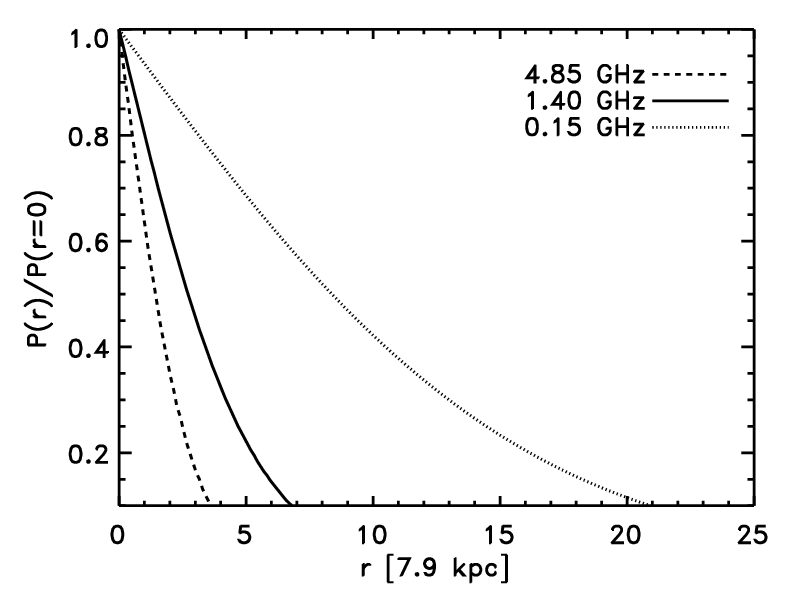}
     \caption{Example of the downstream profiles of the radio emission at $\nu = 0.15 \ \GHz$ (dashed), $1.4 \ \GHz$ (solid) and $4.85 \ \GHz$ (dashed). The profiles have been normalised to the radio emission at the shock front. The shock properties are: $M=2.3$, $T_{\mathrm{d}} = 1.2 \cdot 10^8 \ \K$ and $B = 0.39 \ \mu \G$. The x-axis is in units of the grid resolution $\dd x = 7.9 \ \kpc$.}
     \label{fig::downstream_prof}
 \end{figure}

 In the following section, we give an overview of the model to compute the radio emission. For more details, we point to the Appendix (see App. \ref{app::synchrotron}). We compute the downstream profile of the synchrotron emission from shocked cells following the approach of \citet{2007MNRAS.375...77H}.  \\
 For simplicity, we assume that the properties of the shock front do not change within the electron cooling time. Hence, both the magnetic field strength and the downstream temperature at the shock determine the downstream profile. This assumption is crude as both quantities might affect the shape of the profile. Yet in most cases, the downstream width at frequencies, $\nuobs > 1.0 \ \GHz$, is smaller than both the physical resolution of the simulation and the effective resolution of the MHD scheme\footnote{See Sec. \ref{sec::conclusion} for an elaborated discussion on the resolution of the Dedner cleaning procedure, which is $\sim 32 \ \kpc$.} (see Fig. \ref{fig::downstream_prof}). Hence, the constant shock properties are a valid assumption for frequencies above $1.0  \ \GHz$. \\
 For each shocked cell, we compute the downstream profile of the radio emission as a function of the distance to the shock front. The emission per volume at a distance $x$ is the convolution of the electron spectrum $n_{\mathrm{E}}(\tau, x)$ and the modified Bessel function $F(1/\tau^2)$: 
 \begin{align}
     \frac{\dd P}{\dd V \dd \nu }(x) &= C_{\mathrm{R}}  \int_{0}^{E_{\max}} n_{\mathrm{E}}(\tau, x) F\left(\frac{1}{\tau^2}\right) \dd \tau .  \label{eq::dPdVdv} 
 \end{align}
The electron spectrum at a distance $x$ form the shock is thus:
 \begin{align} 
   \begin{split}
      n_{\mathrm{E}}&(E, x) =  \frac{n_{\mathrm{e}} C_{\mathrm{spec}} }{\me c^2} \left(\frac{E}{\me c^2}\right)^{-s}   \\
      &\times \left[1-\left(\frac{\me c^2}{E_{\max}} + C_{\mathrm{cool}}\frac{x}{v_d}\right)\frac{E}{\me c^2} \right]^{s-2}.
     \end{split} \label{eq::ne}
 \end{align}
 In the equations above, $\tau$ depends on the electron energy. Electrons are considered to be suprathermal if their energy is above $E_{\min} = 10 k_{\mathrm{b}} T$, using the Boltzmann constant $k_{\mathrm{b}}$, and they are accelerated to a finite energy $E_{\max}$. Therefore, we only compute the spectrum if $EC_{\mathrm{cool}} x /v_d / \me / c^2< 1 - E/E_{\max}$, with the downstream velocity $v_d$ and the cooling constant $C_{\mathrm{cool}}$ (see App. \ref{app::synchrotron}).  \\
 The normalisation of the spectrum $C_{\mathrm{spec}}$ depends on the acceleration efficiency $\xi_{\mathrm{e}}$. In the framework of DSA, it is difficult to reach the observed radio luminosities using weak Mach numbers to inject electrons from the thermal pool, as the resulting particle distributions are steep. In order to produce an observable radio relic, we assume an acceleration efficiency of $\xi_{\mathrm{e}} = 0.02$. Furthermore, we include re-acceleration following the approach of \citet{2015MNRAS.451.2198V}. Hence, we assume that shock accelerates a distribution of pre-existing ``fossil'' cosmic-ray electrons \citep[e.g.][]{2013MNRAS.435.1061P}, which boosts the emission of a factor $\approx 100$, for the considered Mach number regime, compared to the single injection case in \citet[][]{2007MNRAS.375...77H}. In the downstream, we compute the radio emission on nodes that lie along the shock normal and that have a fixed distance of $\dd x = 1 \ \kpc$. We assign the emission to the grid cell that hosts the node, assuming that the shock surface matches the cell surface. The emission volume at each node is thus $V_{\rm em} = (7.9 \ \kpc)^2 \ \cdot 1 \ \kpc \approx 62.4 \ \kpc^3$. In Fig. \ref{fig::downstream_prof}, we give an example of the downstream radio emission profile produced by the same shock, $M = 2.3$, at three different frequencies. As the width of the downstream profile changes with frequency, the total volume of the relic increases with decreasing frequency.\\
 We use the same algorithm to compute the parallel, $\Ppara$, and perpendicular, $\Pperp$, component of the radio emission as:
 \begin{align}
     \frac{\dd \Ppara}{\dd V \dd \nu }(x) &= C_{\mathrm{R}} \int_{0}^{E_{\max}} n_{\mathrm{E}}(\tau, x) \left[F\left(\frac{1}{\tau^2}\right)-G\left(\frac{1}{\tau^2}\right)\right]\dd \tau \label{eq::ppara} \\
     \frac{\dd \Pperp}{\dd V \dd \nu }(x) &= C_{\mathrm{R}} \int_{0}^{E_{\max}} n_{\mathrm{E}}(\tau, x) \left[F\left(\frac{1}{\tau^2}\right)+G\left(\frac{1}{\tau^2}\right)\right]\dd \tau \label{eq::pperp} .
 \end{align}
 The functions $F(x)$ and $G(x)$ depend on the modified Bessel functions (see App. \ref{app::synchrotron}). \\
 After applying the above described algorithm, multiple grid cells contain radio emission. For our analysis, we will only use cells, which fulfil two conditions: the emissivity on the three-dimensional grid has to be larger than zero and the cell has to contribute to a bright pixel in the radio map with a luminosity above $4 \cdot 10^{24} \ \erg/ \sek / \Hz$, i.e. the sum of the emission all cells which contribute to the pixel has to be larger than that value. To describe the characteristic physical properties of the radio relic and to provide a sense of what would dominate observations, we introduce, in addition to the arithmetic mean, an radio-weighted average. Using the radio luminosity $P_i$, we computed the radio-weighted average of a quantity $Q$ as 
 \begin{align}
  \langle Q \rangle_{\nuobs} = \sum_i (Q_i P_{i} \left(\nuobs\right) ) / \sum_i P_{i} \left(\nuobs\right). \label{eq::radio_weighting}
 \end{align}
 The sum is taken across all radio-emitting cells that lie along the LoS. As the total emission volume changes with frequency, the radio-weighted average $\langle Q \rangle_{\nuobs}$ also depends on the observing frequency $\nuobs$. In App. \ref{app::angles}, we describe how we take the cyclic property of angles into account when averaging angles.
 \subsection{Polarisation}\label{ssec::polarisation}
 We compute the integrated polarised emission of our radio relic following the formalism of \citet{1966MNRAS.133...67B}:
\begin{align}
  \Pburn(\lambda^2) = \frac{\sum\limits_{\mathrm{los}} P_{\mathrm{tot}} \Pi \exp \left(2i \left(\epsilon_{\mathrm{int}} + \RM \lambda^2 \right) \right) \dd s}{\sum\limits_{\mathrm{los}} P_{\mathrm{tot}} \dd s} \label{eq::burn},
\end{align}
 using the emission per volume in each simulation cell $P_{\mathrm{tot}}$, the intrinsic degree of polarisation $\Pi$ and the intrinsic angle of polarisation $\epsilon_{\mathrm{int}}$. $\RM  \lambda^2$ accounts for Faraday rotation. The intrinsic degree of polarisation, $\Pi$ at observing frequency $\nuobs$, is computed using the parallel and perpendicular component of the radio emission, Eq. \ref{eq::ppara} and \ref{eq::pperp} \citep[e.g.][]{rybickiandlightman}:
\begin{align}
 \Pi = \frac{P_{\parallel}-P_{\perp}}{P_{\parallel}+P_{\perp}} . \label{eq::intrinsic_dop}
\end{align}
 The intrinsic angle of polarisation, $\epsilon_{\mathrm{int}}$, is computed with respect to the horizontal axis of the projected maps. Each simulation cell can be considered to be filled with a uniform magnetic field and, in this case, the intrinsic angle of polarisation is perpendicular to the direction of the projected magnetic field.  If the emission is going through a magnetised medium, the intrinsic angle of polarisation is Faraday rotated. In Eq. \ref{eq::burn}, $\RM  \lambda^2$ determines the amount of Faraday rotation. Here, $\lambda$ is the wavelength corresponding to the observation frequency and $\RM$ is the rotation measure (RM) of the ambient medium. The RM at a distance $x$ from the observer is computed as:
\begin{align}
 \RM = 812 \int_{0}^x \frac{n_\mathrm{e}}{10^{-3} \ \cm^{-3}}  \frac{B_{\para}}{\mu\G} \frac{\dd l}{\kpc} \ \left[\mathrm{\frac{rad}{m^2}}\right]. \label{eq::phi_RM}
\end{align}
 The integral is taken along the LoS. $n_{\mathrm{e}}$ and $B_{\para}$ are the thermal electron number density and parallel magnetic field component, respectively, along the LoS. Faraday rotation occurs either outside of the emitting region, \textit{external} Faraday rotation, or inside the source, \textit{internal} Faraday rotation. \\
 The summation of Eq. \ref{eq::burn} provides a complex number, from which the  polarisation angle ($E$-vector) is computed as:
\begin{align}
  \epsilon_{\mathrm{pol}}(\lambda^2) = \frac{1}{2} \arctan \left( \frac{\mathrm{Im}(\Pburn(\lambda^2))}{\mathrm{Re}(\Pburn(\lambda^2))} \right) . \label{eq::beta}
\end{align}
 We additionally compute the angle of the $B$-vector
 \begin{align}
 \beta_{\mathrm{pol}}(\lambda^2) = \epsilon_{\mathrm{pol}}(\lambda^2) - 90^{\circ}. \label{eq::epsilon}
 \end{align}
 We note that the $B$-vector only corresponds to the magnetic field direction in case of a uniform field and without Faraday Rotation. In a more complex situation, it is only a measure for the observed polarisation angle. \\
 Several effects can reduce the degree of polarisation \citep[e.g.][]{1991MNRAS.250..726T,1998MNRAS.299..189S,2011MNRAS.418.2336A}. In an extended source, the polarised emission emitted at the far side will undergo a different amount of Faraday rotation than the one emitted at the near side leading to depolarisation. Also, instrumental effects can cause depolarisation. \textit{Beam depolarisation} occurs if different polarisation orientations lie within the same telescope beam and hence annul each other partly. Other instrumental effects, such as \textit{bandwidth depolarisation}, can depolarise the emission. We did not include any of those effects and we point for more information to textbooks such as \citet{2015gimf.book.....K} and references therein. 
 \section{Results}\label{sec::results}
 \subsection{Simulated Emission from Radio Relics}\label{ssec::sample}
\begin{figure}
\includegraphics[width=0.49\textwidth]{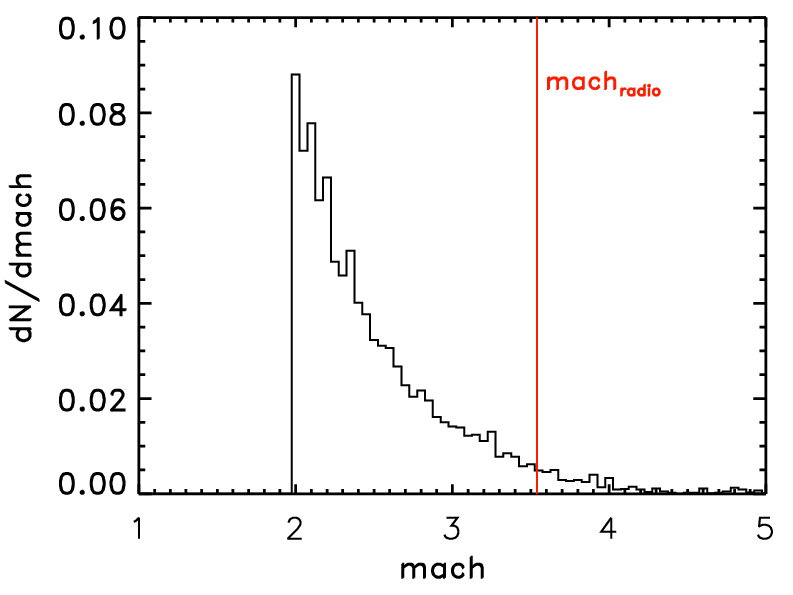} \\
\includegraphics[width=0.49\textwidth]{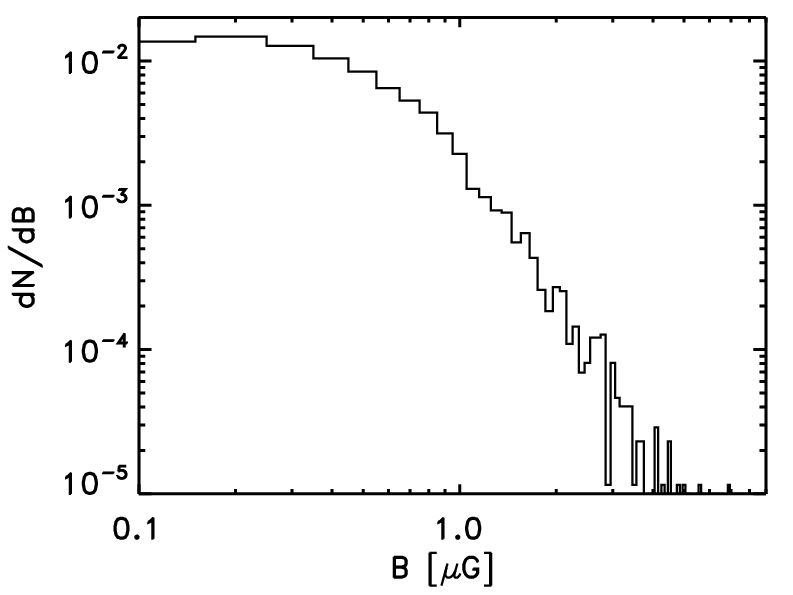} 
  \caption{Volumetric distributions of Mach number (top) and magnetic field strengths (bottom) of radio-emitting cells at $1.4 \ \GHz$. The distributions have been normalised to the number of cells that emit in radio. The red line in the top plot marks the Mach number computed from the spectral index between $1.4 \ \GHz$ and $0.15 \ \GHz$.}
  \label{fig::b_hist}
\end{figure}
\begin{figure*}
  \includegraphics[width = \textwidth]{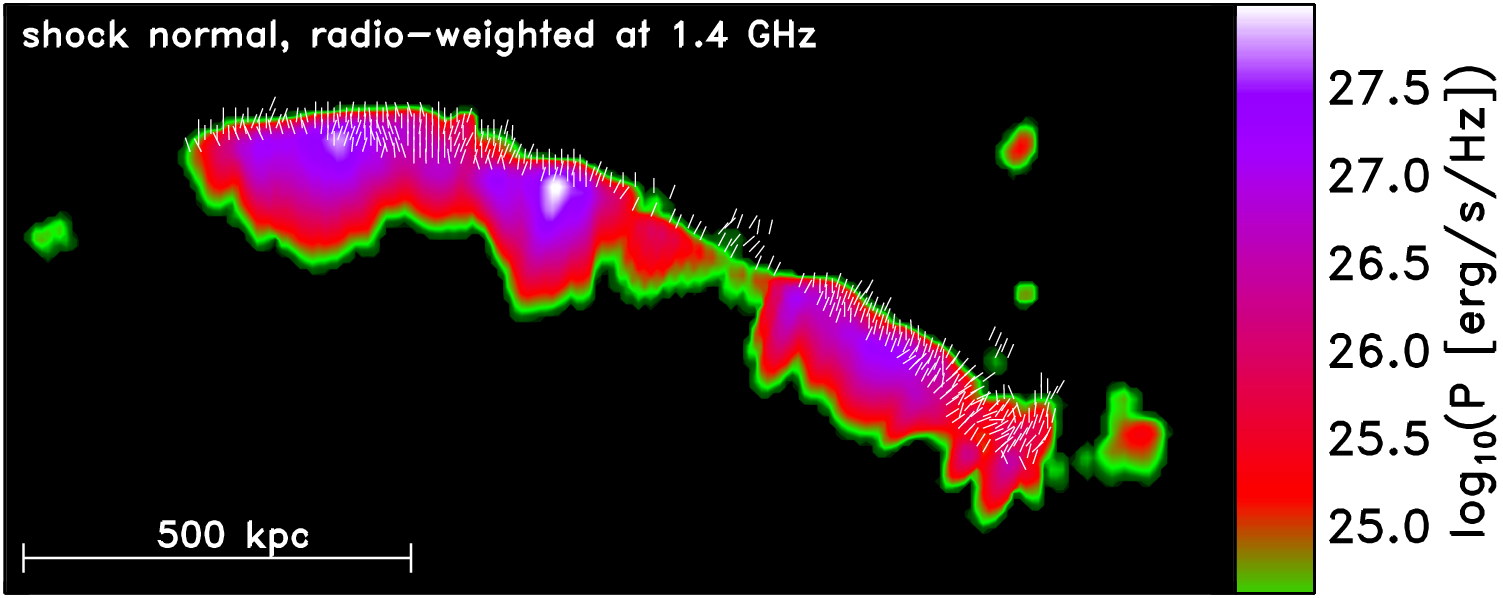} \\
  \includegraphics[width = \textwidth]{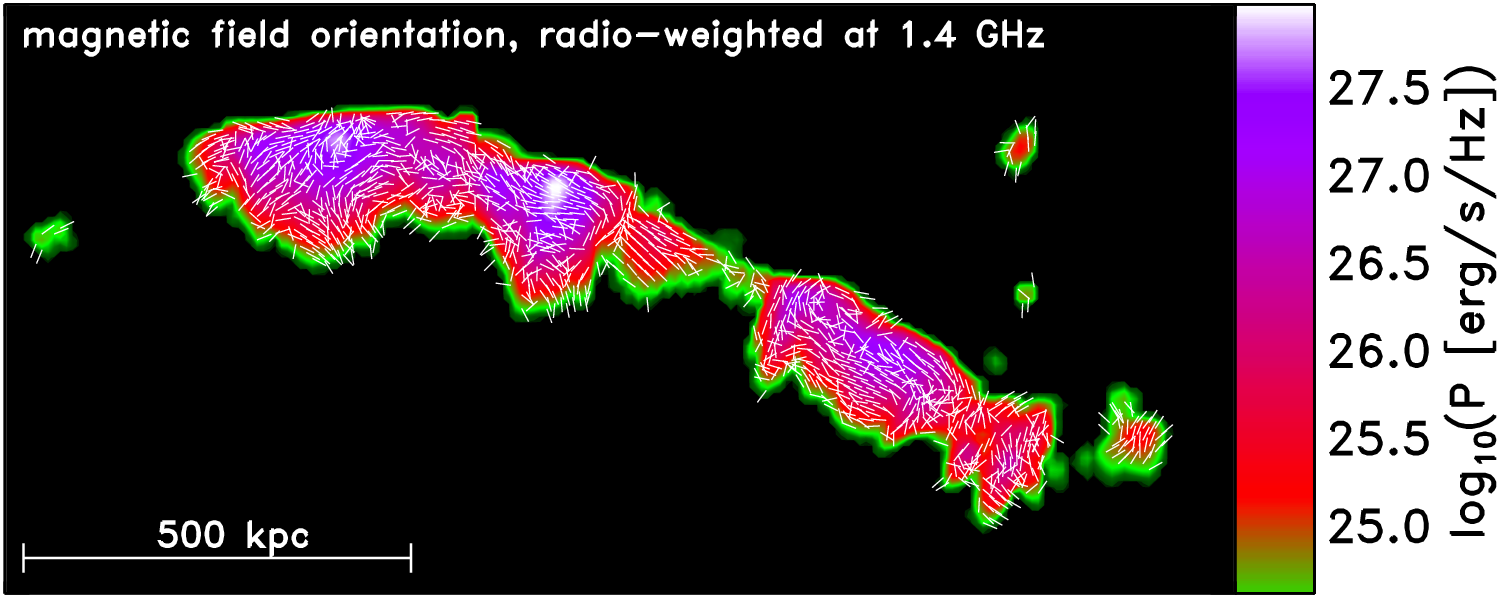}  
  \caption{Radio luminosity (colour) overlayed with the radio-weighted projected shock normal (top) and the radio-weighted orientation of the physical magnetic field (bottom). The vectors have been normalised to unity.}
  \label{fig::burn_true}
\end{figure*}
\begin{figure}   
  \includegraphics[width = 0.49\textwidth]{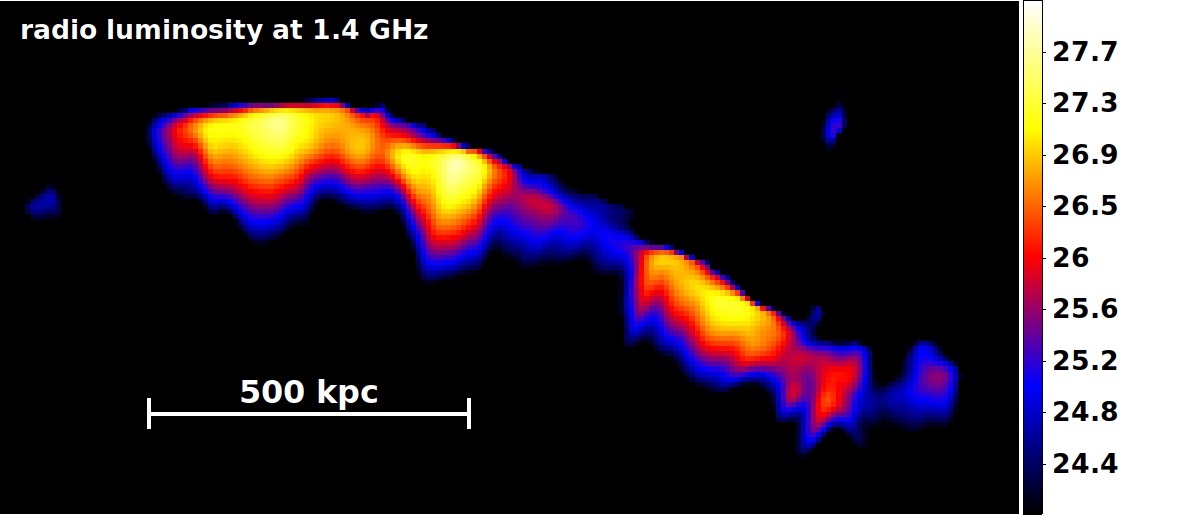} \\
  \includegraphics[width = 0.49\textwidth]{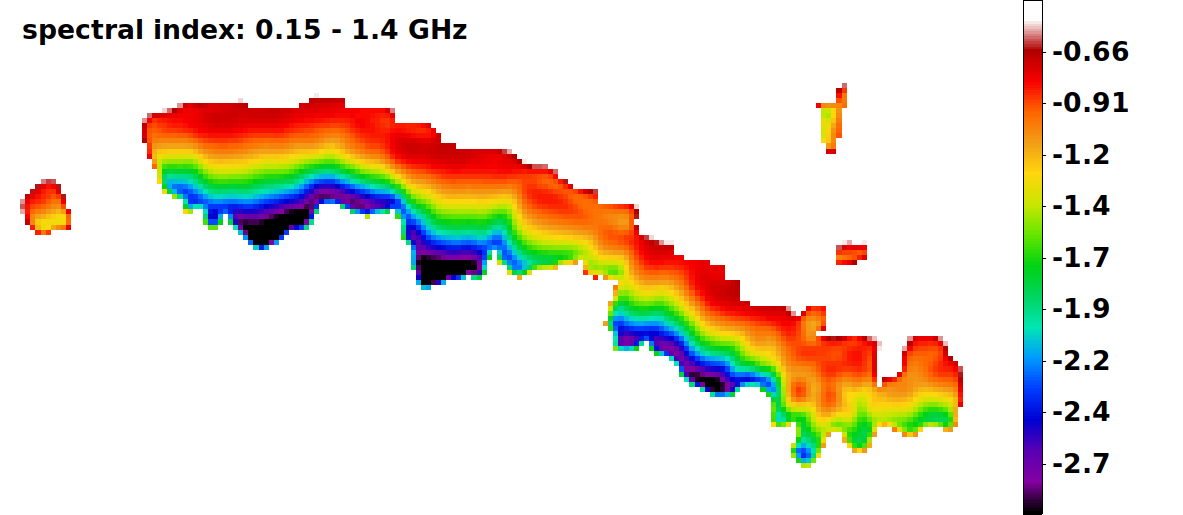}\\
  \includegraphics[width = 0.49\textwidth]{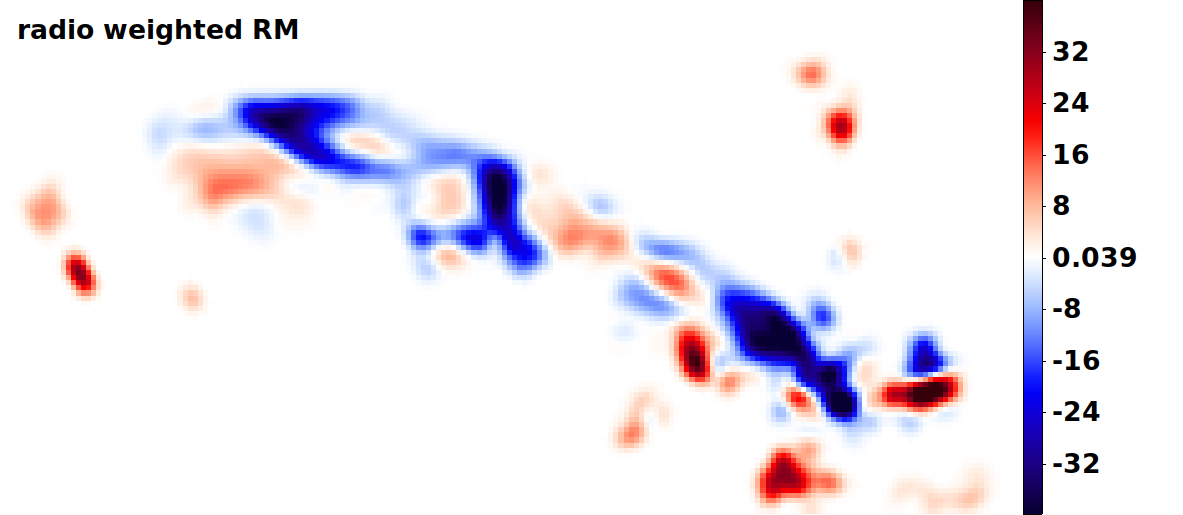} \\
  \includegraphics[width = 0.49\textwidth]{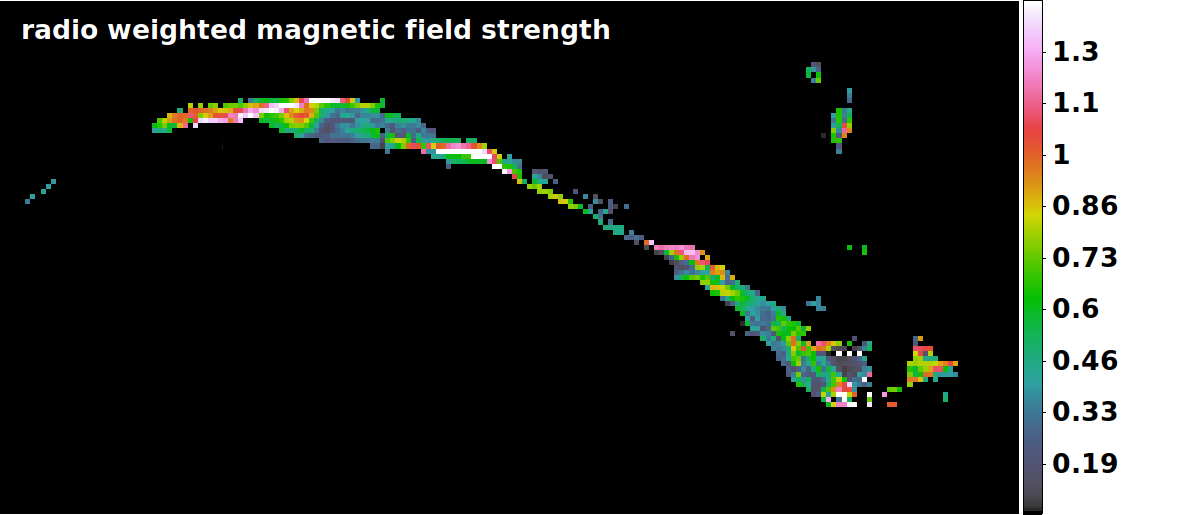} \\
  \includegraphics[width = 0.49\textwidth]{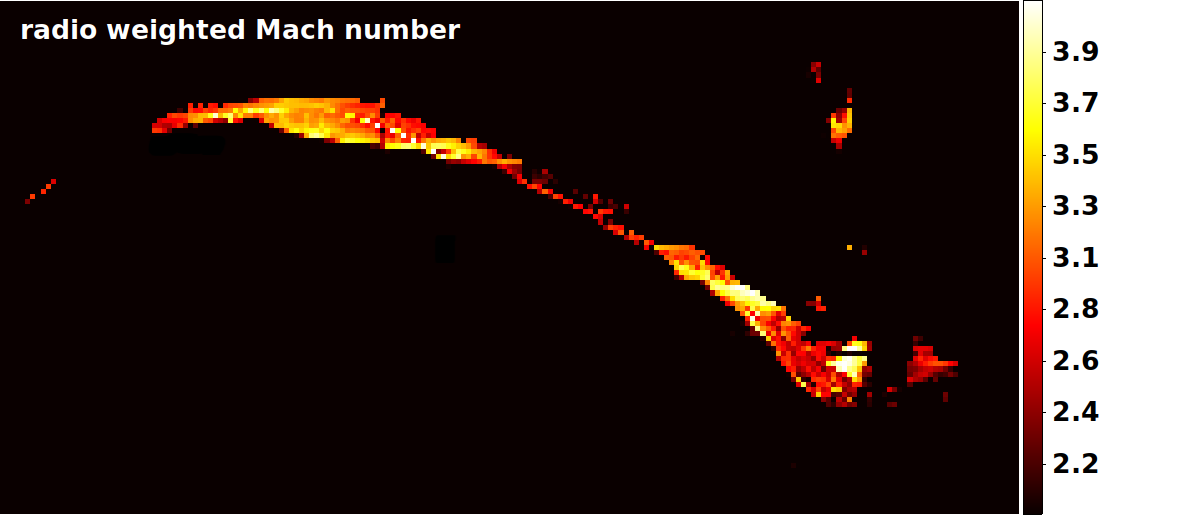} 
  \caption{Physical quantities at the relic's position. From top to bottom: radio luminosity in $\log_{10}(\erg/\sek/\Hz)$, spectral index between $0.15 \ \GHz$ and $1.4 \ \GHz$, radio-weighted RM in $\rad/\m^2$ as well as the radio-weighted magnetic field strength in $\mu \G$ and radio-weighted Mach number at the shock surface. In all cases, the radio weighting was done at $1.4 \ \GHz$. (A coloured version is available in the online article.)}
  \label{fig::rel_xy_ingredients}
\end{figure}

 So far, only \citet{2013ApJ...765...21S} and \citet{2017MNRAS.464.4448W} have studied the properties of magnetic fields in radio relics in cosmological simulations. The simulation in the present work has an unprecedented numerical resolution, that is necessary to evolve a small-scale dynamo \citep[see][for details]{2018MNRAS.474.1672V}, giving us a plus for computing the polarised emission in radio relics. \\
 In this work, we focus on the re-simulation of a $\sim 10^{15} \ \Msun$ galaxy cluster that undergoes a major merger at $z \approx 0.2$, producing two powerful shock waves. In Fig. \ref{fig::rtr}, we plot the gas density overlaid with radio emission contours at $\nuobs = 1.4 \ \GHz$ projected along the three different axes of the simulation box. The edge-on view shows two prominent, large-scale shock waves that produce radio emission on a $\sim \ \Mpc$ scale. For our analysis, we focused on the larger and brighter one of the two (see blue arrows in Fig.  \ref{fig::rtr}). For the analysis, we only included those cells that lie in the area of $(1580 \times 790) \ \kpc^2$ around the relic (i.e. see top panel in Fig. \ref{fig::rel_xy_ingredients}). If the same relic is observed along the two other orthogonal LoS, then the radio emission becomes more extended and dimmer. In the remainder of this paper, we will refer to the relic as it is observed in the three different projections: "edge-on", "face-on" and "side-on" as labelled in Fig. \ref{fig::rtr}. In the following, we will compare the typical quantities of the simulated relic to  observations. \\ 
 The radio relic in our simulation is at a distance of $\sim 1.7 \ \Mpc$ to the centre of mass of the cluster  (see Tab. \ref{tab::relics}).  Its apparent morphology varies significantly with the projection. When observed edge-on, it has a small surface that is thin and elongated, while when observed both face-on and side-on, they show much larger surfaces and more filamentary structures. The radio power of the relics is of the order of $\sim 10^{30} \ \erg/\sek/\Hz$ at $1.4 \ \GHz$. The small discrepancies between the different projections are due to the emission in front or behind the relic along the LoS. The simulated relic lies below the mass-luminosity relation derived by \citet{2014MNRAS.444.3130D}, but one has to take into account that this relation has been derived using a much more powerful class of double radio relics and that it might be biased towards brighter objects due to the sensitivity of radio telescopes. Still, the simulated relic is fainter than most observed single radio relics. This could be a consequence of either the low Mach number or the significantly low magnetic field strength at the relics position. \\
 For several observed radio relics, the Mach number derived from radio observations is larger than the one derived from X-rays \citep[e.g.][and references therein]{2015ApJ...812...49H,2018MNRAS.478.2218H}. This discrepancy could be related to projection effects \citep{2015ApJ...812...49H} or systematic errors in the X-ray observations \citep[][]{2017A&A...600A.100A}. Hence, we calculated the Mach number distribution across the shocked cells that have $M \ge 2$ (see top panel in Fig. \ref{fig::b_hist}). Additionally, we computed the integrated radio spectral index\footnote{$ \alpha_{\mathrm{R}} = \frac{\log_{10}\left(P_{1.4}/P_{0.15}\right)}{\log_{10}\left(\nu_{1.4}/\nu_{0.15}\right)}$} for each projection, and the corresponding Mach number\footnote{$M = \frac{2\alpha_{\mathrm{R}}+2}{2\alpha_{\mathrm{R}}-2}$} between $1.4 \ \GHz$ and $0.15 \ \GHz$. The spectral index obtained for each projection is $\alpha_{\mathrm{R}} \approx -1.17$ corresponding to Mach numbers of $M \approx 3.5$. Therefore, the Mach number derived from the spectral index is much larger than most of the Mach numbers in the simulation (see red line in Fig. \ref{fig::b_hist}). These findings are in agreement with \citet{2018ApJ...857...26H}, where they found that Mach numbers derived from radio observations are biased towards larger values because the shock acceleration efficiency strongly depends on the Mach number.\\
 The average Mach number and average magnetic field strength are given in Tab. \ref{tab::relics}, where we additionally show their radio-weighted averages (see Eq. \ref{eq::radio_weighting}). In the remainder of the paper, we will discuss the other properties of the simulated relic and we will focus mainly on the edge-on view. We choose the edge-on view because the relic morphology is similar to the assumed ``prototype'' of radio relics and  because the direction of the shock normal is well defined (see. Fig. \ref{fig::burn_true}). Throughout the work, we will highlight differences and similarities with the other two LoS. 
\subsection{Distribution of magnetic fields}\label{ssec::magneticfields}
 The maps of projected gas density, temperature and magnetic field strength at the relics position are shown in Fig. \ref{fig::rel_environment}. We can observe that the upstream density and temperature are very regular at the top left edge of the relic. Yet at the lower right edge, a sub-clump is falling into the cluster, causing this sector to be disturbed. The upstream magnetic field shows a similar topology, whilst we observe some small-scale fluctuations in the downstream. Finally, we can observe that the shock front is not uniform as it is highlighted better in the temperature map. \\
 Fig. \ref{fig::b_hist} shows the magnetic field strength's distribution of the radio-emitting cells. Most of the radio-emitting cells have magnetic field values that are smaller than $ 1 \ \mu \G$ and only a few cells have magnetic field strengths above $ 2 \ \mu \G$. The average magnetic field strength producing the radio relic is in the range of $0.2-0.4 \ \mu \G$, while the radio weighted magnetic field strength is $\sim 1.05 \ \mu \G$. These magnetic field values are smaller than estimations at the position of observed relics \citep[see Fig. 14 in][]{2013MNRAS.433.3208B}. This occurs in our simulation since probably the small-scale dynamo did not have sufficient time to evolve and amplify the magnetic field at the relic's position. \\
 In Fig. \ref{fig::burn_true}, we show the shock normal as well as the radio-weighted (see Eq. \ref{eq::radio_weighting}) orientation of the projected magnetic field at $1.4 \ \GHz$  in order to get a sense of the magnetic field behaviour. The radio-weighted magnetic field direction changes rapidly across the length of the relic. We can observe that the magnetic field aligns with the shock surface in several regions,  while it is perpendicular to the shock surface in others. \\
 We also compute the magnetic power spectrum and its correlation length, i.e. the outer scale of the spectrum, in the $(7.9 \times 200 \ \kpc)^3$ sub-box centred on the relic. We obtained a correlation length of $\sim  248.8 \ \kpc$, by Fast Fourier-transforming the three-dimensional magnetic field and by fitting its power spectrum with the functional form derived in \citet{2019arXiv190311052D}. We refer the reader to the Appendix \ref{app:1} for more information on the magnetic spectrum.\\
 In Fig. \ref{fig::rel_xy_ingredients}, we show different properties of the radio relic itself. First, we show the radio emission at $1.4 \ \GHz$ and the spectral index map computed between $0.15 \ \GHz$ and $1.4 \ \GHz$. The spectral index is a computed by superimposing different radio downstream profiles and hence, the ageing of the spectral index is slower than expected from Fig. \ref{fig::downstream_prof}. Next, we show the radio-weighted (see Eq. \ref{eq::radio_weighting}) RM at the relic's position, the radio-weighted magnetic field strength and the radio-weighted Mach number at the shock front. The radio emission varies significantly across the relic, and two bright patches are visible. The latter connect via a fainter and filamentary bridge and they are located in the regions of strong Mach numbers and of large magnetic field values. We can also observe that the RM is not uniform across the relic due to the varying magnetic field strengths and densities along the LoS. 
 \subsection{Distributions of Rotation Measures} \label{ssec::rm}
\begin{figure*}
  \includegraphics[width = 0.33\textwidth]{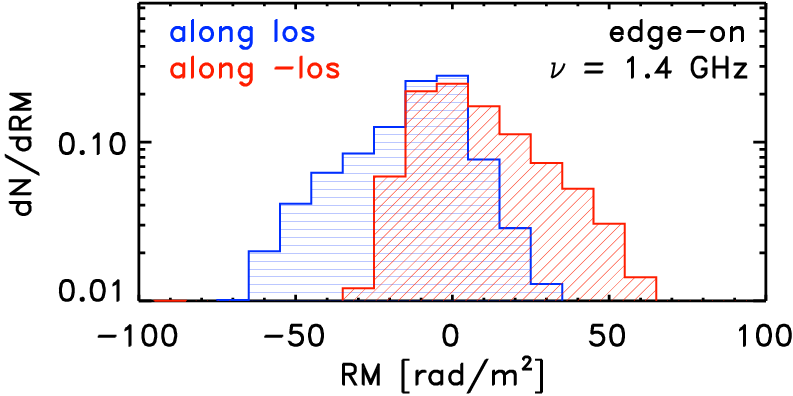} 
  \includegraphics[width = 0.33\textwidth]{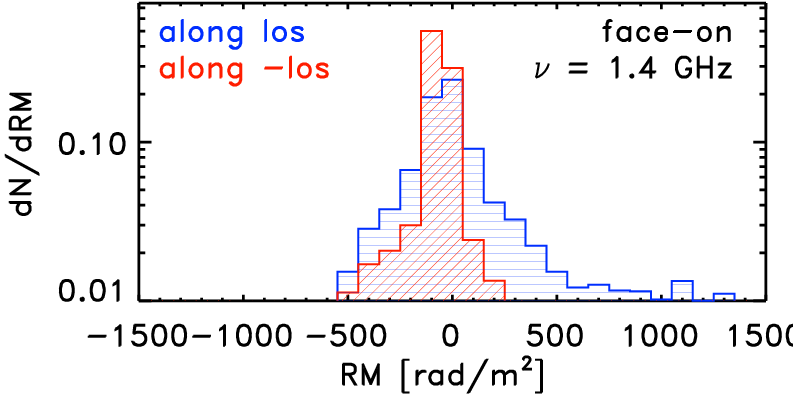} 
  \includegraphics[width = 0.33\textwidth]{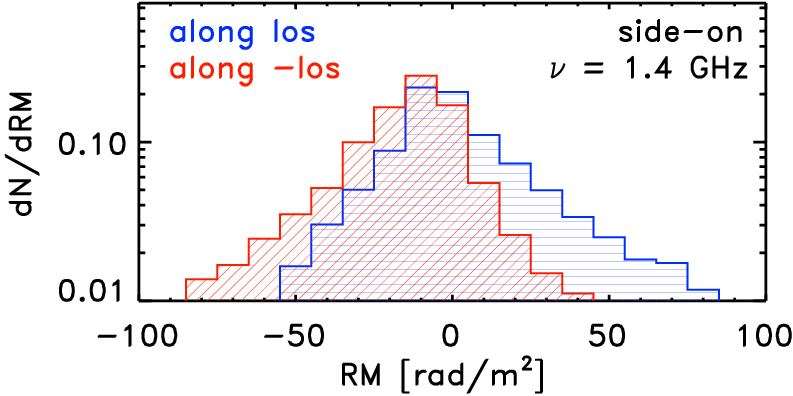} \\
  \includegraphics[width = 0.33\textwidth]{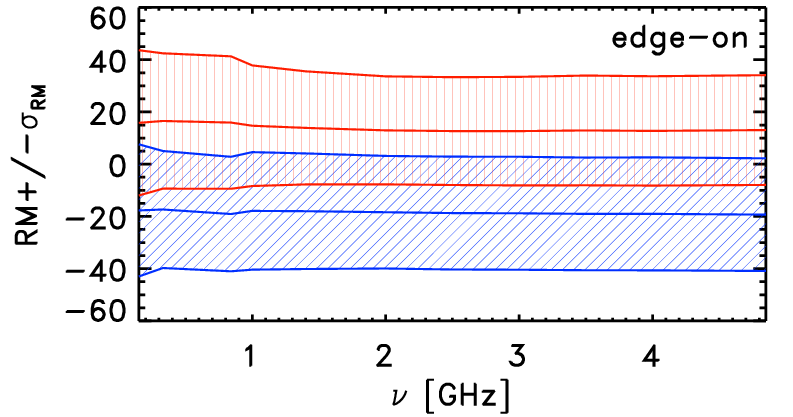} 
  \includegraphics[width = 0.33\textwidth]{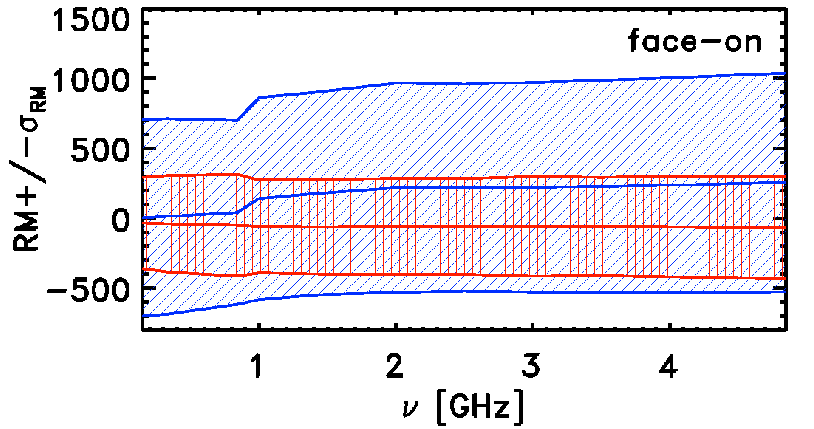} 
  \includegraphics[width = 0.33\textwidth]{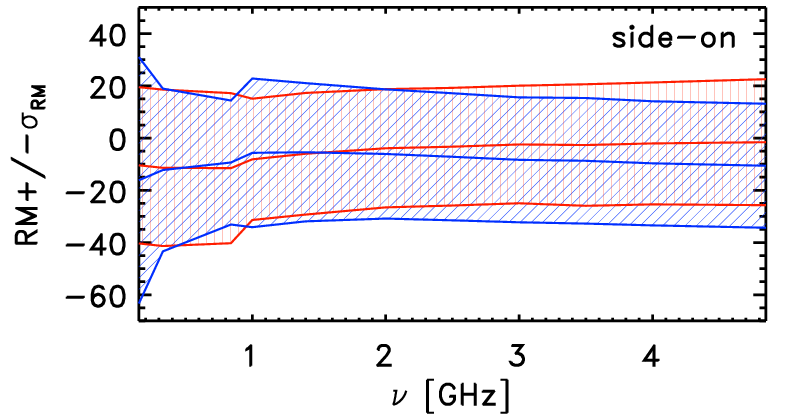} 
  \caption{Normalised distributions of the Rotation Measures in the radio-emitting cells for the edge-on, face-on and side-on view (left to right). We only show the distributions where $\dd N / \dd \RM >1 \ \%$. The top row shows the distributions of the radio-emitting cells at $1.4 \ \GHz$ for each projection (the different range of the x-axis in the central panel should be noticed). The bottom panels show the average values (in $[\rad/\m^2]$) plus/minus the one standard-deviation of the distributions as a function of frequency. The colours match the colours of the distribution: blue (dash-dotted) along the LoS and red (dotted) along the LoS but in opposite direction.   (A coloured version is available in the online article.)}
  \label{fig::rm_hist}
\end{figure*}
 In the presence of magnetic fields, the polarisation vectors undergo Faraday rotation. The amount of rotation depends on both the strength and distribution of the RM along the LoS. As the RM distribution is most likely not uniform for an extended source, the polarisation vectors will rotate differently and therefore this could enhance depolarisation (see Sec. \ref{ssec::polarisation}). In the near future, RM measurements will become more precise due to the increasing resolution of telescopes and the application of RM Synthesis \citep[e.g.][]{2015aska.confE..92J,2015aska.confE..95B,2005A&A...441.1217B}, and polarisation studies will improve. \\
 In the following, we will describe the RM distribution in our simulation in greater detail. We obtain two different RM distributions for each projection, since Eq. \ref{eq::burn} can be integrated from different sides of the computational domain. We will refer to these cases as "along LoS" and "along  -LoS". Including the case without Faraday rotation ($\RM = 0$  in Eq. \ref{eq::burn}), gives a total of nine (independent) test cases to study the polarisation of the radio relic. \\
 First, we compute the distribution of RM found in the radio-emitting cells. As the surface and the depth of the emitting region change with frequency, the RM distribution depends on the observing frequency as well. In the top panels of Fig. \ref{fig::rm_hist}, we show the normalised RM distributions at $1.4 \ \GHz$, where we consider only values of $\dd N / \dd \RM >1 \ \%$. In the bottom panels of Fig. \ref{fig::rm_hist}, we plot the average and standard deviation of each distribution for different frequencies. \\
 The standard deviation for the edge-on and side-on view are a few tens $\rad/\m^2$ and they remain fairly constant with frequency. On the other hand, when seen face-on, the standard deviation is one order-of-magnitude higher. In this face-on view, the distributions are much broader and show extended tails with values above $\pm 200 \ \rad/\m^2$ for at least one viewing direction. The relic is behind the cluster in this view (see Fig. \ref{fig::rtr}), and so the emission must pass through a longer magnetised region. \\
 The average values of the different distributions remain almost constant at high frequencies, while their variation becomes larger at low frequencies. These variations are still within one standard deviation. The fact that we obtain less variation at high frequencies might have a numerical nature as the cooling regions at these frequencies are under-resolved. The effective spatial resolution of any hydro-MHD scheme ($32 \ \kpc$ in our simulation) is coarser than the nominal one, and therefore, we cannot account for any variation of the RM on smaller scales that become more important at higher frequencies.  \\
 Summarising the RM analysis, we obtain standard deviations of the RM distributions of a few $ 10 \ \rad/\m^2$ in most cases of the simulation, and values between $\sim 300 - 800 \ \rad/\m^2$ only if the radio emission originates from behind the cluster. \\
 In the following, we want to examine whether or not the RM in our simulation agrees with the RM observed in the ICM that can be obtained through the polarisation analysis of background sources: \citet{2012arXiv1209.1438H} compiled a RM catalogue of extragalactic sources located at redshift $0<z<5.3$. They found that the variance of the RM distributions does not change with redshift and they estimated a standard deviation of $\sigma_{\RM} \approx 23 \ \rad/\m^2$;  \citet{2016A&A...596A..22B} estimated the RM dispersion in regions located more than $1 \ \Mpc$ away from the cluster centre of a large sample of galaxy clusters, finding a value of $\approx (57 \pm 6) \ \rad/\m^2$; \citet{2004rcfg.proc...51J} studied the RM of the north-western relic in Abell 3667 and found that locally the values can be $\sim -165.1 \ \rad / \m^3$ and $\sim 98.2 \ \rad / \m^2$. Yet, higher resolution studies of the RM distribution showed that the dispersion can be smaller at the relic; \citet{2013MNRAS.433.3208B} measured a dispersion of $\sim 5.2 \ \rad/\m^2$ for the Coma cluster. Consequently, we conclude that the RM and its variation produced in our simulation agrees with current observations. Yet for higher magnetic field values, we would need a corresponding smaller correlation length in order to recover the same $\sigma_{\RM}$ (the correlation lengths of the magnetic field in the whole simulation box are $\sim 250-300 \ \kpc$). \\
 We only have used the $\sigma_{\RM}$ of either the cluster periphery or the entire cluster, and any contribution of the intergalatic medium (e.g. integrating over $\sim800 \ \Mpc$ for $z \approx 0.2$) would have to be added. We do not include them here as these contributions are small and they would not change our results: \citet{2010ApJ...723..476A,2011ApJ...738..134A} obtained a contribution of about $\sim 1 \ \rad/\m^2$ for filaments and an RM saturation of $\sim 7-8 \ \rad/\m^2$ for $z \ge 1$. Furthermore, they showed that the main contributors to the RM are galaxy clusters. These results were also confirmed by \citet{2018MNRAS.480.3907V}, who also argued that intergalactic RM contribution should be even smaller in the case of purely astrophysical magnetic seed fields. \\
 As a final caution, one should notice that the RM distributions measured here are neither Gaussian nor symmetric (see Fig. \ref{fig::rm_hist}). Yet, several works \citep[e.g.][]{1966MNRAS.133...67B,1991MNRAS.250..726T} assume Gaussian or symmetric RM distributions which can lead to systematic effects in the inference of magnetic fields. 
 \begin{figure*}  
  \includegraphics[width = \textwidth]{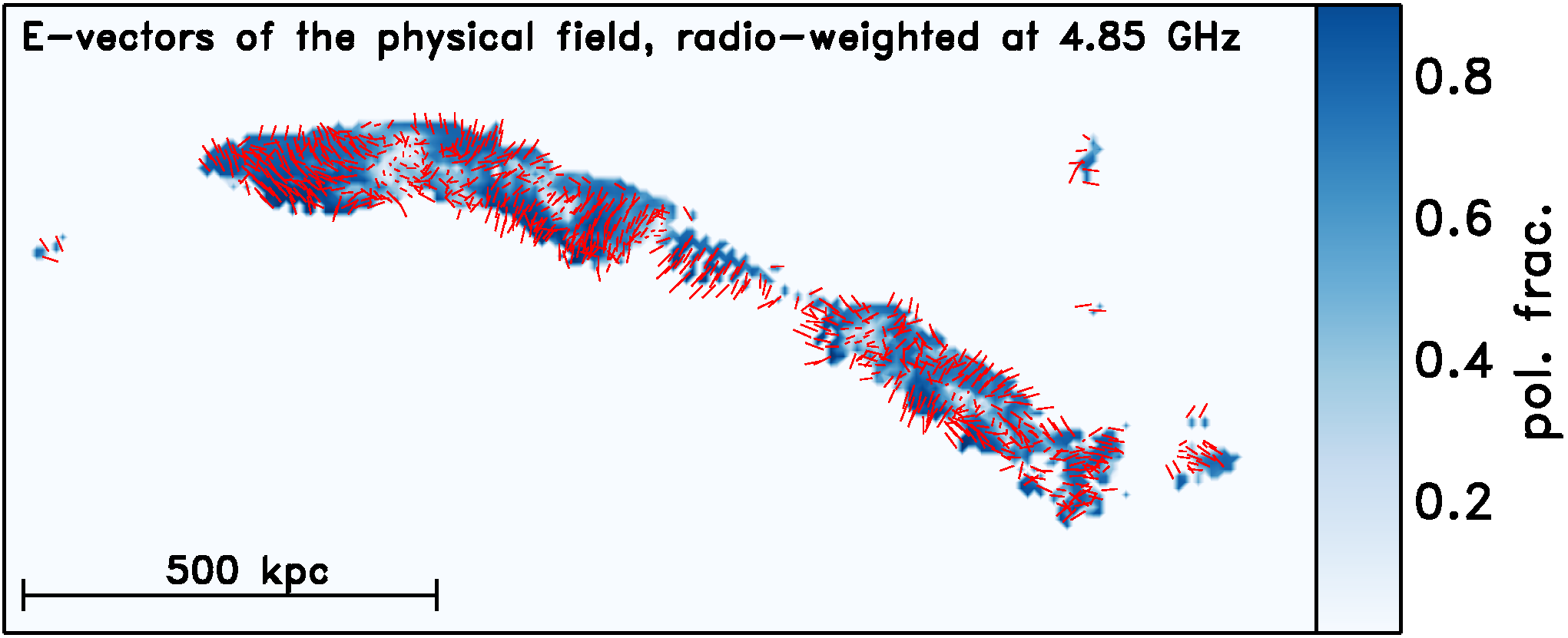} \\
  \includegraphics[width = \textwidth]{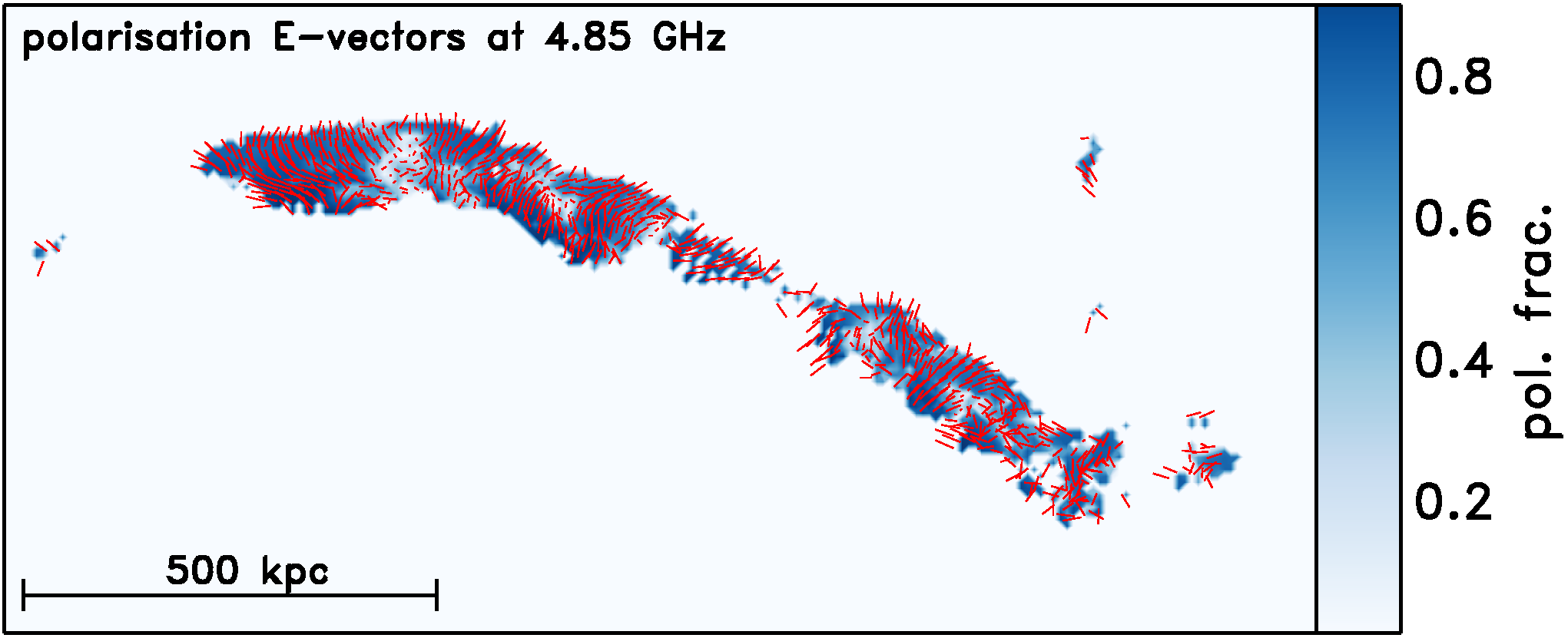}\\
    \includegraphics[width = \textwidth]{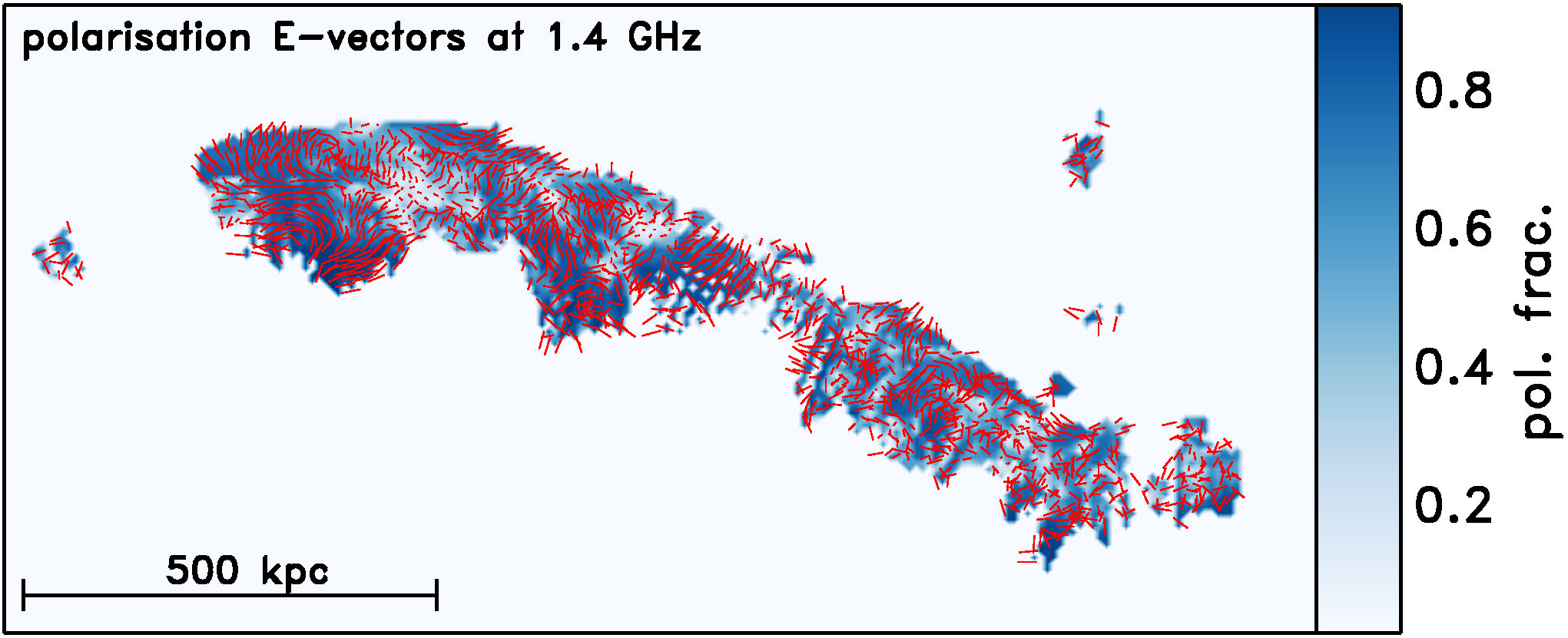} 
  \caption{Projected fractional polarisation (colour) overlayed with the $E$-Vectors (white vectors). The top row shows the radio-weighted $E$-vectors of the physical field at $\nu =  4.85  \ \GHz$. The middle row and the last row show the polarisation $E$-vectors at $\nu = 4.85 \ \& \ 1.4  \ \GHz$. The vectors have been normalised to unity. (A coloured version is available in the online article.)}
  \label{fig::burn_polVec_E_xy}
 \end{figure*}
 \subsection{The intrinsic Properties of the polarised Emission}\label{ssec::properties}
\begin{figure}
  \includegraphics[width = 0.49\textwidth]{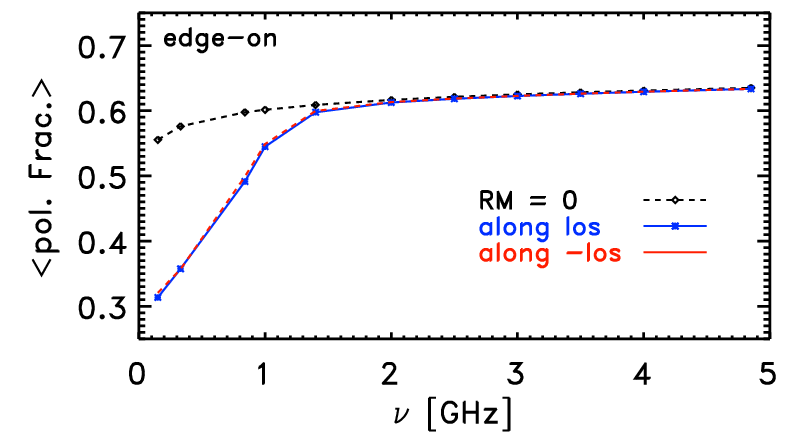} \\
  \includegraphics[width = 0.49\textwidth]{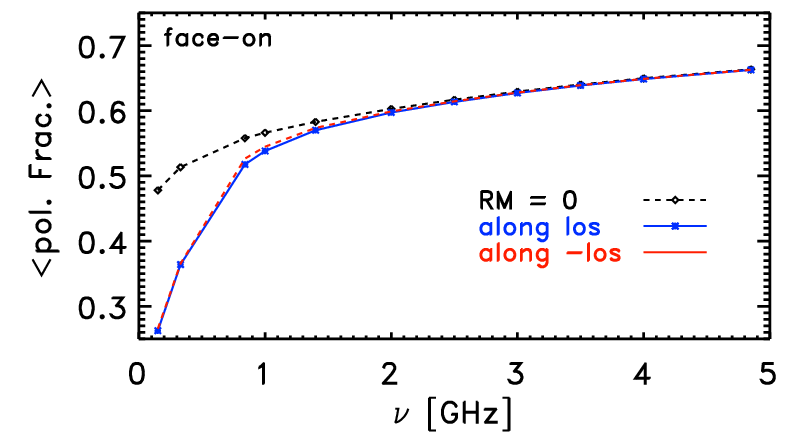} \\
  \includegraphics[width = 0.49\textwidth]{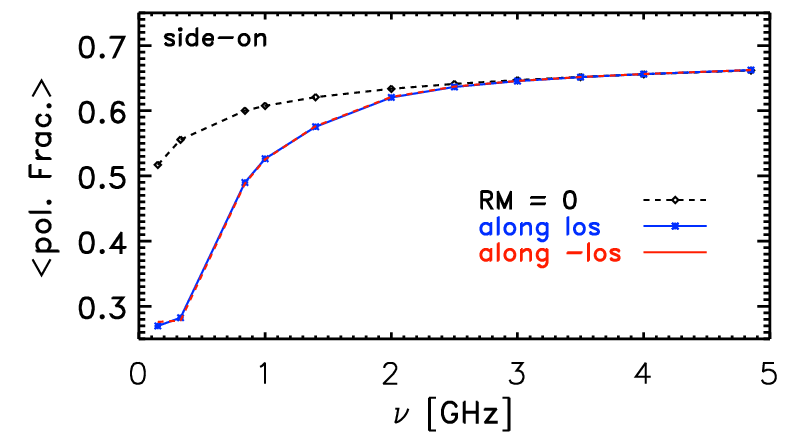} 
  \caption{Radio-weighted degree of polarisation. The different panels display the results for the different projections: edge-on, face-on and side-on (top to bottom). The different lines give the different RM models. (A coloured version is available in the online article.)}
  \label{fig::dop_vs_freq_paper}
\end{figure}
 We computed the integrated polarised emission (Eq. \ref{eq::burn}) and the corresponding $E$-vector (Eq. \ref{eq::epsilon}). As an example, we plot the fractional polarisation overlayed with the corresponding $E$-vectors for two different observing frequencies in the last two panels of Fig. \ref{fig::burn_polVec_E_xy}. Additionally, we show the radio-weighted, at $4.85 \ \GHz$, orientation of the physical electric wave to visualise its characteristic orientation. Locally, the $E$-vectors have the same orientation, but they do not show alignments on scales above $500 \ \kpc$. Also, there is no strong correlation between the direction of the $E$-vectors and the shock normal (compare with Fig. \ref{fig::burn_true}). The orientation of the $E$-vectors seen edge-on starts to randomise locally for $\nu < 1.4 \ \GHz$. On the other hand, at $\nu = 4.85 \ \GHz$ the effect of Faraday rotation is small and we observe a orientation similar to the case without Faraday rotation. Along each LoS, a few bright cells dominate the radio emission. Hence, their intrinsic degrees of polarisation determine the polarisation fraction which does not significantly decrease with frequency. In these maps, we do not apply any observational cut and, hence, the extent of the polarisation fraction into the downstream mirrors the extent of the surface brightness which is larger for lower frequencies. Since the intrinsic degree of polarisation increases for steeper spectra, the polarisation fraction becomes larger in the downstream. \\
 We do not show the $E$-vectors if the relic is seen along the opposite direction of the LoS, as the pattern is similar. In both the face-on and side-on view, the orientation of $E$-vectors is similar to the edge-on view. However, their morphology is more complex and cannot be directly related to the true shock normal. \\
 As observations tend to pick up the brighter parts of the relic, we computed the radio-weighted average degree of polarisation for frequencies in the range from $150 \ \MHz$ to $4.85 \ \GHz$. In Fig. \ref{fig::dop_vs_freq_paper}, we plot the degrees of polarisation versus frequency. We also include the control cases $\RM = 0$ that reflect the intrinsic degree of polarisation. We note that we computed the  RM on scales of the grid resolution and any higher amplitude small-scale structures in the Faraday depth, due to the finite resolution of our simulation, are neglected. Therefore, the degree of polarisation reflects depolarisation by internal Faraday rotation, while external Faraday rotation only rotates the polarisation vectors without causing any additional depolarisation. \\
 Independent of the projection and the $\RM$-selection, the degree of polarisation is always larger at high frequencies, i.e. $\sim 0.64$ seen edge-on and $\sim 0.67$ seen face-on and side-on, and decreases for lower frequencies. The degree of polarisation drops to $\sim 0.31$ seen edge-on and $\sim 0.26-0.28$ seen face-on and side-on at $150 \ \MHz$ for the cases that include Faraday Rotation. Without Faraday Rotation, the degree of polarisation is significantly higher at $150 \ \MHz$, i.e. $\ge 48 \ \%$. The degree of polarisation for the different test cases including RM converges with the control case, i.e. $\RM = 0$, above $1.0 \ \GHz - 2.0 \ \GHz$. \\ 
 One would expect that the face-on projection (which features the highest values of $\RM$) would show the lowest degree of polarisation and the slowest convergence to the control case of $\RM = 0$ due to the $\exp \left(-\RM^2 \lambda^4\right)$ factor. Nevertheless, this is not the case because the extent of the relic along the LoS is much shorter than in the other to cases, i.e. $\sim 160 \ \kpc$ instead of $\geq 400 \ \kpc$. Therefore, the face-on projection effectively probes fewer polarisation vectors and depolarisation since the internal Faraday rotation is small. \\
 In all three control cases, i.e. $\RM = 0$, the degree of polarisation decreases at lower frequencies.  The downstream width of the relic is larger at low frequencies and, hence, polarisation is also probed at a larger distance to the shock front. 
 The farther into the downstream, the more magnetic fluctuations we encounter, (see Fig. \ref{fig::rel_environment}) and as a consequence, the orientation of the intrinsic angle of polarisation is more random. This is also illustrated at low frequencies. \\
\begin{table*}\centering
  \begin{tabular}{l||c|c|c|c|c|c}
   Telescope &   $\nu \ [\GHz]$ & $d_{\mathrm{beam}} \ [\mathrm{arcsec}]$ & $\sigma_{\mathrm{noise}} \ [\mu \mathrm{J} / \mathrm{beam}]$ & $I_T \ [\mu \mathrm{J} / \mathrm{arcsec}^2]$ & pol. Frac. $[\%]$ at $z=0.4/0.1$ & pol. Frac. $[\%]$ at $z=0.5/0.2$ \\ \hline \hline
   LOFAR        & 0.15 & 25  & 500    & 2.1  & 5.7  & 4.8 \\
   VLA          & 1.4  & 5   & 6     & 0.64  & 50.4 & 48.6 \\
   VLA          & 1.4  & 7   & 8     & 0.43  & 47.2 & 45.1 \\
   VLA          & 1.4  & 11  & 12    & 0.26  & 39.6 & 38.8 \\
   VLA          & 1.4  & 16  & 18    & 0.18  & 31.2 & 31.9 \\
   VLA          & 1.4  & 25  & 26    & 0.11  & 26.0 & 29.1 \\
   Effelsberg   & 4.85 & 159 & 800   & 0.083 & 16.1 & 8.6
  \end{tabular}
  \caption{Fiducial parameters for our mock observations. The first two columns give the name of the telescope and the observing frequency. The third, fourth and fifth column provide the beam size, thermal noise and estimated detection threshold respectively. The last two columns give the observed polarisation fraction of the relic is placed at $z = 0.4$ ($z = 0.1$ for Effelsberg) and $z = 0.5$ ($z = 0.2$ for Effelsberg) respectively. The parameters have been taken from Tab. 1 in \citet{2017A&A...598A.104S}, Tab 2. in \citet{2018ApJ...852...65R} and Tab. 3 in \citet{2017A&A...600A..18K}.}
  \label{tab::telescopes}
\end{table*}
\begin{figure*}
  \includegraphics[width = 0.975\textwidth]{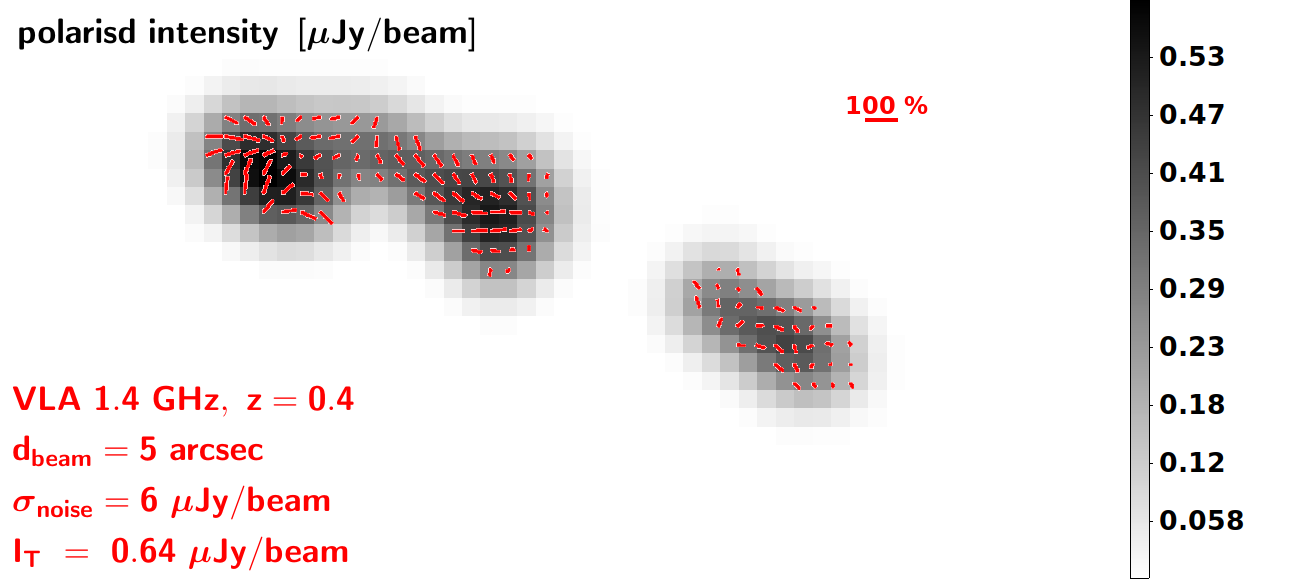}\\ 
  \includegraphics[width = 0.975\textwidth]{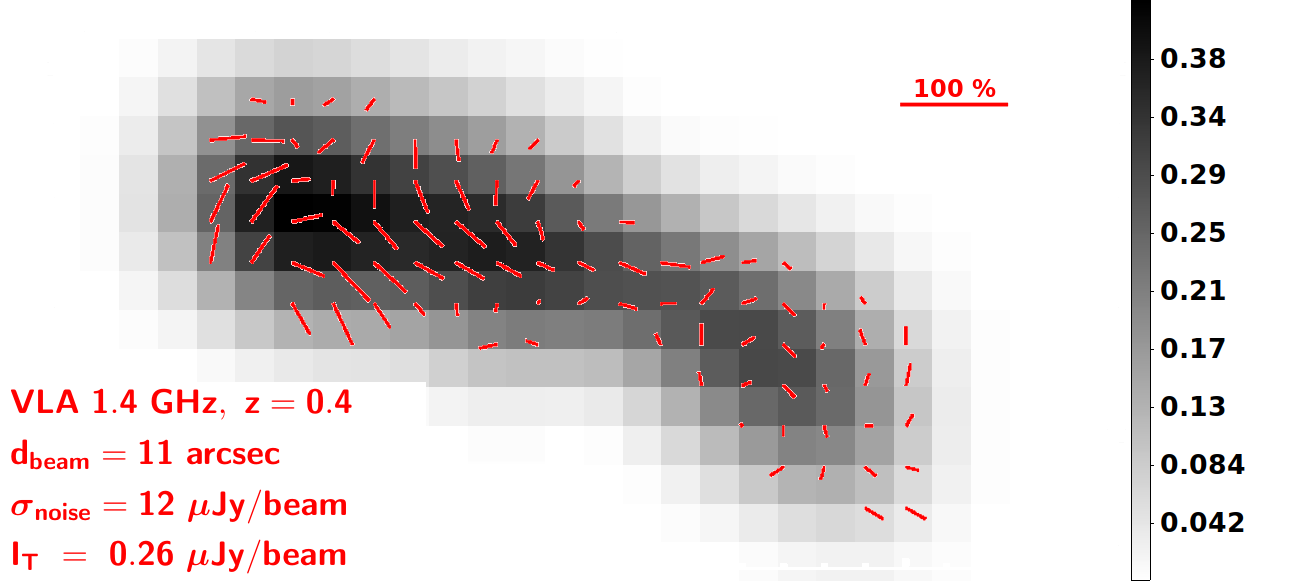}\\ 
  \includegraphics[width = 0.975\textwidth]{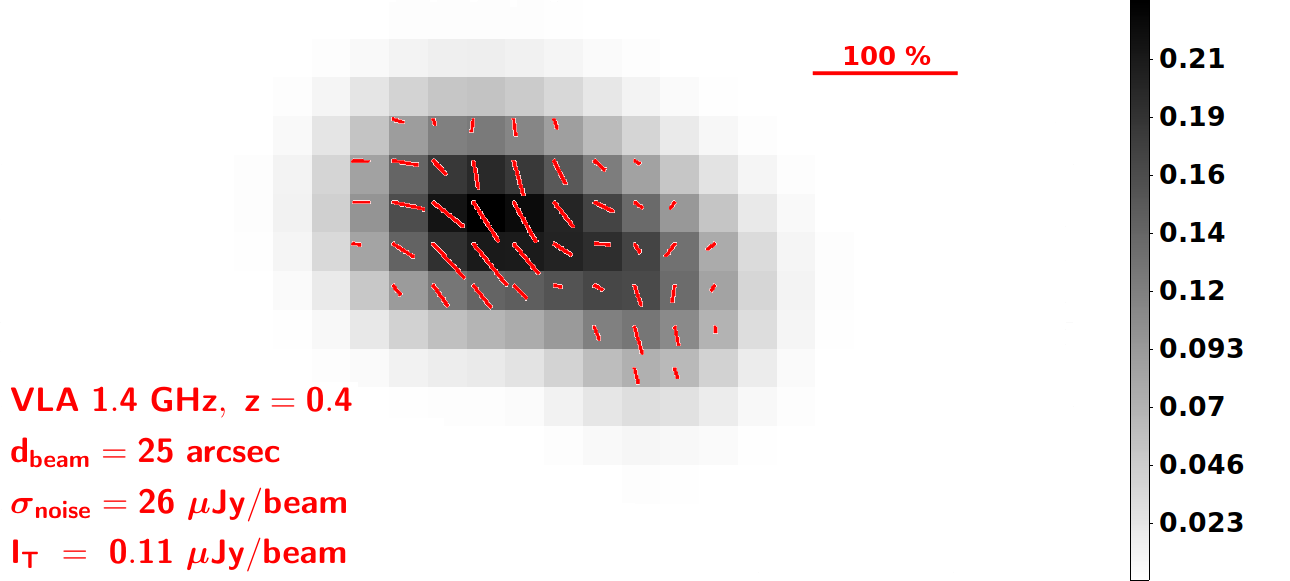} 
  \caption{VLA mock observations, at $1.4 \ \GHz$ and using beam sizes of $5 \ \arcsec$, $11 \ \arcsec$ and $25 \ \arcsec$, of the relic placed at $z = 0.4$. The polarised intensity (color) is overlayed with the $E$-vectors (red vectors). A reference of $100 \ \%$ polarisation is shown in the top right. (A coloured version is available in the online article.)}
  \label{fig::telescope_maps}
\end{figure*} 
 In summary, we obtain an average degree of polarisation $\geq 50 \ \%$ for  $\nu \geq 1.0-1.4 \ \GHz$ and it is significantly larger than what it is obtained in observations  (see Tab. \ref{tab::relics_obs}). On the other hand, our results are in line with previous \enzo  \ simulations by  \citet{2013ApJ...765...21S}, who measured a maximum degree of polarisation $\sim 75\ \%$. We reiterate that the degree of polarisation only accounts for depolarisation due to internal Faraday rotation and, thus, including the beam and/or bandwidth depolarisation can decrease it.
 \subsection{Synthetic Radio Observations}\label{ssec::telecsope}
\begin{figure}
  \includegraphics[width = 0.49\textwidth]{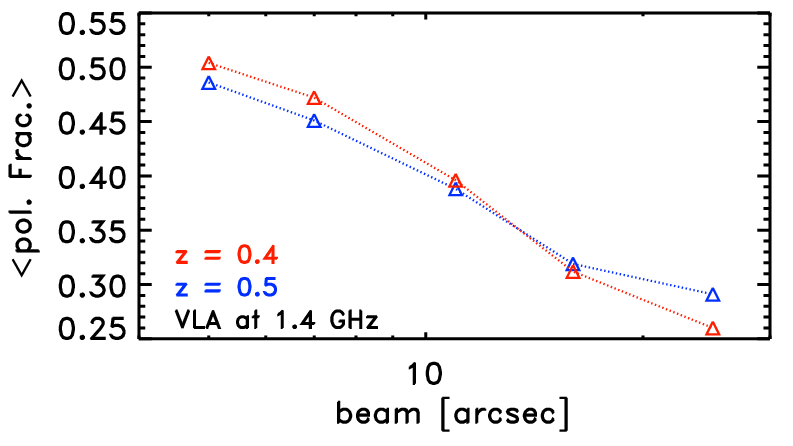} \\
  \includegraphics[width = 0.49\textwidth]{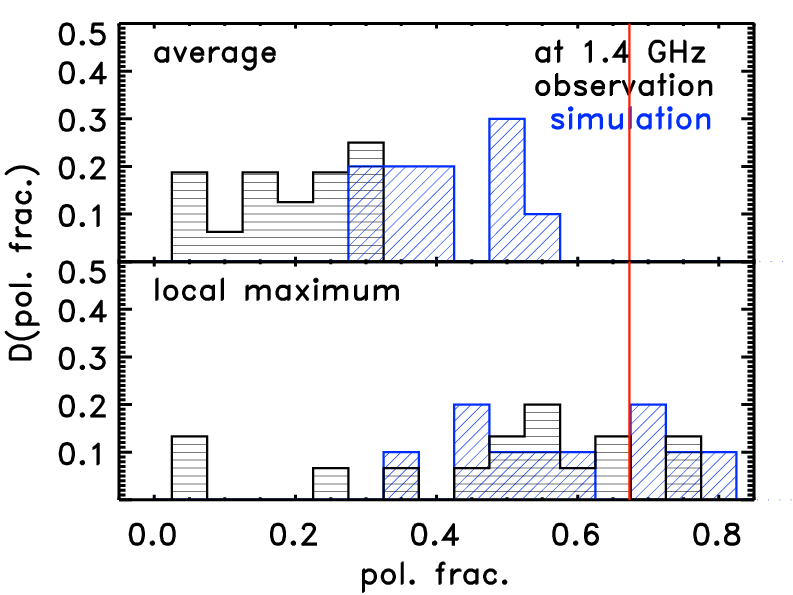} 
  \caption{Top: radio-weighted degree of polarisation as a function of the beam size. The two lines show the results for the VLA mock observations at redshift $z = 0.4$ (red line) and $z = 0.5$ (blue line). Bottom: distributions of the polarisation fraction measured by observations (black) and in the simulation (blue) $1.4 \ \GHz$. The top distributions compare the average degree of polarisation across the relic, while the bottom distributions compare the local maximum degree recorded in each relic. The red line marks the polarisation fraction $75 \ \%$ computed by \protect{\citet{2013ApJ...765...21S}} (A coloured version is available in the online article.)}
  \label{fig::depolarisation}
\end{figure}
\begin{figure*}
  \includegraphics[width = \textwidth]{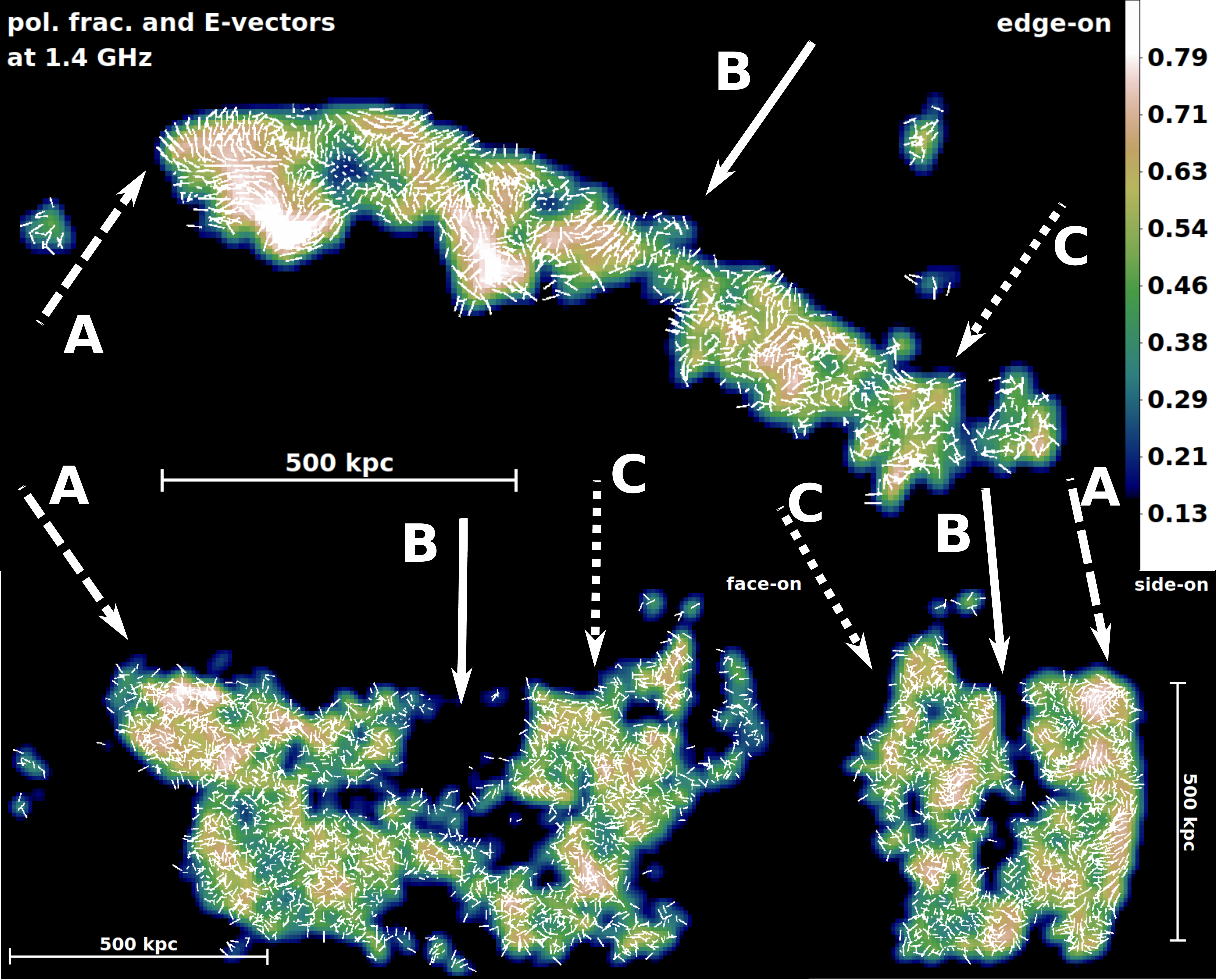}
  \caption{Polarisation fraction (colour) overlayed with the corresponding $E$-vectors (white vectors). The three different panels show the relic seen in different projections. The letters highlight regions that have remarkable similarities with structures observed in real radio relics: (A) alignment of $E$-vectors, (B) a bridge and (C) a depolarised "brush". The vectors have been normalised to unity. (A coloured version is available in the online article.)}
  \label{fig::sim_to_obs}
\end{figure*}
 In order to compare with real radio observations, we performed synthetic observations of our cluster with the Effelsberg telescope, the VLA and LOFAR-HBA. After locating our relic at different redshifts and converting the emitted power into the physical frame, we convolved the luminosity and the integrated polarised emission that were previously finely sampled (see Eq. \ref{eq::dPdVdv} and \ref{eq::burn} respectively), to the proper resolution/beam size of the assumed telescope configurations, $d_{\mathrm{beam}}$. The detection threshold,$I_T$, is computed as a flux density level of three times the noise per beam, $\sigma_{\mathrm{noise}}$: $I_T = 3 \times \sigma_{\mathrm{noise}} / (1.1333 \cdot d_{\mathrm{beam}})^2$.\\
 We considered the fiducial parameters shown in Tab. \ref{tab::telescopes} which were taken from recent papers. In particular, we use the parameters given in Tab. 1 of \citet{2017A&A...598A.104S} for the direction-independent \textit{LOFAR Two Meter Sky Survey} for the LOFAR-HBA mock observation. We use the parameters in Tab. 2 of \citet{2018ApJ...852...65R} for the VLA mock observation and parameters in Tab. 3 of \citet{2017A&A...600A..18K}
 for the Effelsberg mock observations. In the case of the Effelsberg mock observation, we place the relic at $z = 0.1 \ \mathrm{and} \ 0.2$, as the large beam would cover more than the entire simulation box for larger redshifts. On the other hand, for the LOFAR-HBA and VLA mock observation, we have to use $z = 0.4 \ \mathrm{and} \ 0.5$, as they would have a higher physical resolution than our reconstructed mesh at smaller redshifts. Only a few polarisation studies using LOFAR at $\nu \sim 0.15 \ \GHz$ have been published so far \citep[e.g.][]{2019A&A...622A..16O}, but none concerning radio relics. Therefore,  we only include these mock observations as a sanity test for our implementation. In the following, we neglect any sort of bandwidth depolarisation \citep[e.g.][]{1996A&AS..117..137H,1996A&AS..117..149S}. We provide the radio-weighted degrees of polarisation for each mock observation in Tab. \ref{tab::telescopes}. \\
 The degree of polarisation of the Effelsberg observation is $ 16.1 \ \%$ at $z = 0.1$ and $ 8.6 \ \%$ at $z = 0.2$. Beam depolarisation occurring in the large telescope beam, $\sim (159 \ \arcsec)^2$ corresponding to $(64 \ \kpc)^2$ and $(112 \ \kpc)^2$ at $z = 0.1$ and $0.2$, respectively, produce these small values. These results agree with the $5 \ \%$ polarisation in Abell 1612 ($z \approx 0.172$) reported by \citet{2017A&A...600A..18K} who used the Effelsberg telescope to study the polarisation of four radio relics at $4.85$ and $8.35 \ \GHz$. Yet, our results are at odds with their measurements of $36 \ \%$ and $15 \ \%$  for the relics in CIZA J2242.8+5301 ($z \approx 0.192$) and  in 1RXS J0603.3+4214 ($z \approx 0.225$). \\
 Our LOFAR-HBA mock observations at $0.15 \ \GHz$, show a polarisation degree below $6 \ \%$ at both redshifts. We expect that this is a numerical artefact and has no physical meaning. Hence, we do not expect to detect significant polarisation at LOFAR-HBA frequencies. \\
 In Fig. \ref{fig::telescope_maps}, we plot the integrated polarised emission overlayed with the corresponding $E$-vectors for three different VLA mock observations at $z = 0.4$. Furthermore, we show the degree of polarisation for the different VLA configuration in Fig. \ref{fig::depolarisation}. Our results lie in the range of $26 \ \%- 50 \ \%$and they decrease with increasing beam size, which indicates beam depolarisation at both redshifts. \\
 In order to compare with real observations, we compute first the distributions of the average and the local maximum degree of polarisation of all radio relics studied in polarisation at $\nu \approx 1.4 \ \GHz$ (see Tab. \ref{tab::relics_obs}). Then, we compare these to the distribution functions from our VLA mock observations at $\nu = 1.4 \ \GHz$ (see Tab. \ref{tab::telescopes}), and plot the two distribution functions in the bottom panel of Fig. \ref{fig::depolarisation}. The distribution functions (average and local maximum) of the mock observations have lower dispersion and peak at higher values than the corresponding distribution functions of the real observations. Even though our final results mildly agree with polarisation observations, a larger statistical sample of simulated relics is needed to test whether their properties agree with reality.
 \subsection{Morphological Properties}\label{ssec::comparison}
 The observed morphology of the polarised emission is quite diverse (see references in Sec. \ref{ssec::relic_obs}). Hence, our small sample cannot explore the variety given by observations. Our relic does not show an overall global alignment of $E$-vectors as observed in some radio relics. However, it shows local structures that match features observed in radio relics, so it may offer useful hints to interpret observed polarisation structures. In the following, we want to highlight three cases that show qualitative similarities with observed structures. \\
 In Fig. \ref{fig::sim_to_obs}, we show a close-up view of our radio relic at $1.4 \ \GHz$ seen in the three different projections. The arrows point to three different regions that we want to further investigate: a region with locally aligned $E$-vectors (A), a bridge structure (B) and a depolarised ``brush'' (C). In this section, we do not apply any detection threshold or re-binning to a specific beam size as high-frequency radio observation can indeed resolve structures of $8 \ \kpc$ or smaller in radio relics \citep[e.g.][]{2018ApJ...852...65R}. We notice, that the discussed radio structures are significantly larger than the effective resolution of our hydro-MHD scheme. Even if some features appear to be close to the effective resolution ($\sim 32$ kpc), this is only due to the projection along the LoS, while their intrinsic 3-dimensional separation is much larger. 
 \subsubsection*{Bridge Structure}
 The bridge structure connects the two brighter patches of the radio relic seen edge-on. Compared to other regions, this region is very filamentary in the face-on view. In the edge-on view, the thickness of the bridge is about $\sim 150 \ \kpc$. Though, in the face-on view, the filaments' extend is about $\sim 260 \ \kpc$. Hence, the vectors are probing magnetic field structures which are significantly separated. If falling within the same telescope beam, they can get depolarised. The edge-on view of the Mach number (see Fig. \ref{fig::rel_xy_ingredients}) reveals a filamentary shock structure in the bridge region. New high-resolution radio observations of relics show similar complex structures in the form of filaments and threads \citep[e.g][]{2018ApJ...852...65R,2018ApJ...865...24D}. Whether these structures are dominated by magnetic fields, by cosmic-ray driven small-scale instabilities, or by a complex shock morphology, cannot be established at present. The threads of the Mach number in the simulation are a  product of (magneto)-hydrodynamic flows and they are most-likely visible if our simulated relic was observed at a higher resolution. The other parts of the relic could in principle have similar filamentary structures, but they might only appear in the bridge because there is no projection of other emissions on top.
 \subsubsection*{Alignment of $E$-vectors}
 In the top-left region of the radio relic,  the $E$-vectors are locally aligned to each other as seen in most radio relics. In the edge-on view, the relic's extent, along the LoS, is about $\sim 134 \ \kpc$ at the left edge and it increases to $\sim 434 \ \kpc$ at the right edge. In the side-on view, only the $E$-vectors that are close to the right edge (i.e. the shock front), align with each other. Finally, in the face-on view, 
 their orientation is completely randomised by Faraday rotation due to the large $\RM$ of this projection. The position of the region coincides with the position of the regular and laminar pre-shock area where the magnetic field is aligned in front of the relic (see Fig. \ref{fig::rel_environment}).
 \subsubsection*{Depolarised ``Brush''}
 In contrast to the strong alignment of $E$-vectors in region A, we observe that the $E$-vectors in region C are completely misaligned. This region looks similar to the ``brush'' in 1RXS J0603.3+4214 \citep{2012A&A...546A.124V}. Moreover, in both the face-on and the side-on view, the $E$-vectors in this region are randomly orientated. The random orientation in the face-on view can be explained by the large RM. But its nature in the edge-on and the side-on view must be physical since their $\RM$ values are fairly low. We infer that the random orientations are due to the cold sub-clump falling onto the cluster in this region (see Fig. \ref{fig::rel_environment}). The sub-clump stirs both the upstream gas and the upstream magnetic fields causing the magnetic field orientation to be random and the $E$-vectors not to align. \\
 \newline
 In summary, our simulation does not reproduce the morphology observed in relics such  as CIZA J2242.8+5301 or Abell 2744. Whereas locally, we recover structures as have been found in observed radio relics. We conclude that threads and filaments can be very distant to each other along the LoS and hence they can probe magnetic field structures with different orientations. We found that aligned polarisation vectors reside in the region of the shock where the upstream magnetic field is laminar, and that they are orientated randomly in the region with a disturbed upstream magnetic field. 
\begin{figure*}
    \includegraphics[width =0.33\textwidth]{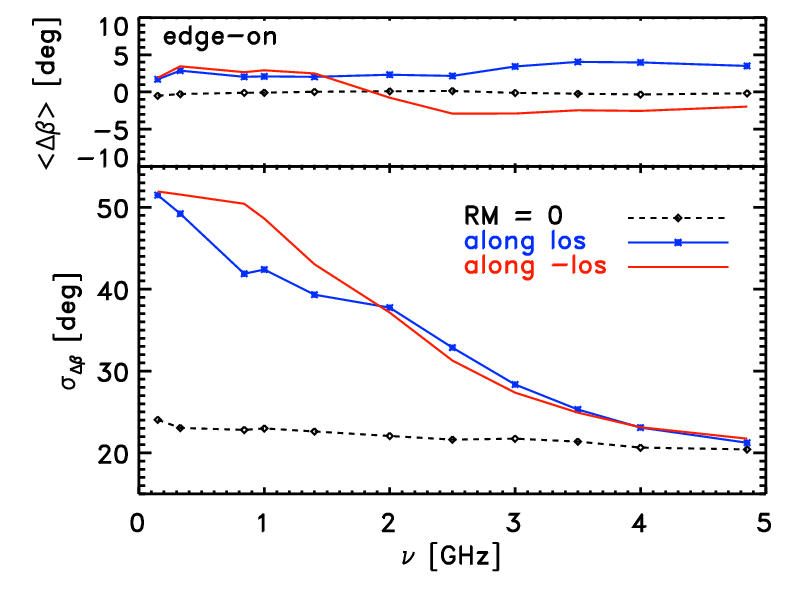}  
    \includegraphics[width =0.33\textwidth]{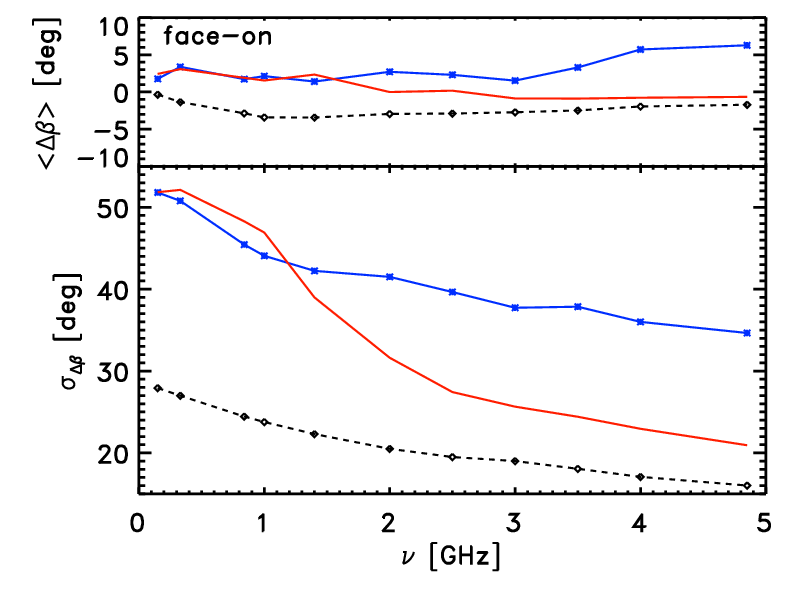}  
    \includegraphics[width =0.33\textwidth]{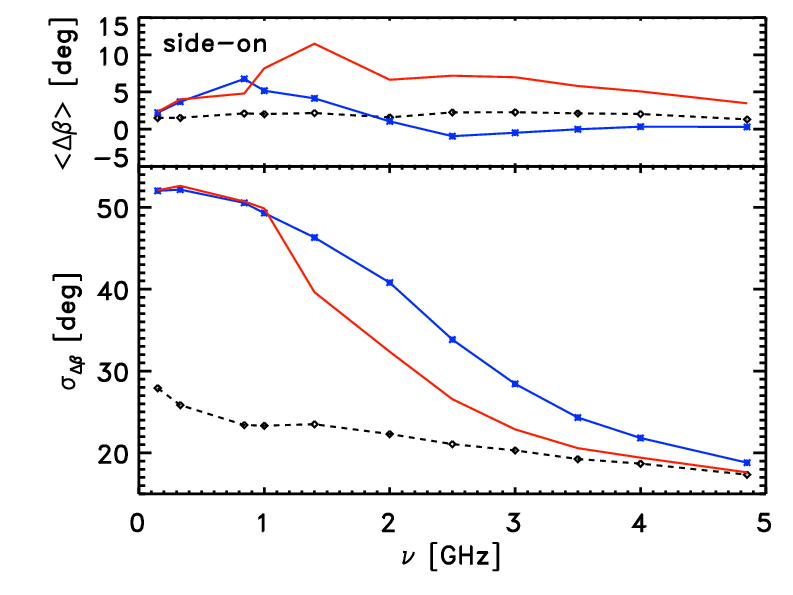} 
  \caption{The plots show the average (top row) and corresponding standard deviation (bottom row) of the  differences between the radio emission weighted magnetic field orientation and the magnetic field orientation derived from polarisation. The black lines (dashed with diamonds) give the cases without RM, while the blue (solid with squares) and red (solid) lines give the cases with RM seen along the two different directions of the LoS. The different panels display the results for the three different projections. (A coloured version is available in the online article.)}
  \label{fig::beta_diff}
\end{figure*}
 \subsection{Magnetic Field Vectors and Polarisation Direction} \label{ssec::magnetic}
 It is possible to derive the magnetic field direction from the observed polarised emission (see Eq. \ref{eq::beta}), provided one accounts for the effect of Faraday rotation. We derived the magnetic field direction for different frequencies between $0.15 \ \GHz$ and $4.85 \ \GHz$ using Eq. \ref{eq::beta}. We subtracted for the effect of Faraday rotation by subtracting $\langle \RM \rangle_{\nuobs} \lambda^2_{\mathrm{obs}}$ from each polarisation angle. In our case, a good proxy for the orientation of the physical magnetic field of the relic is its radio emission weighted orientation, which we computed in the same frequency range (see App. \ref{app::angles}). We computed the differences, $\Delta \beta \in [-90^{\circ}, +90^{\circ}]$, between the corrected polarisation orientation, $B$-vectors and the radio-weighted orientation of the physical field. We also included the control case ($\RM = 0$) as this gives the intrinsic orientation of the polarised emission. Finally, we obtained a distribution of differences across the relic for each frequency and each LoS.\\
  In Fig. \ref{fig::beta_diff}, we plot the mean and the dispersion (i.e. using the standard deviation) of these distributions. Regardless of both frequency and LoS, the mean values fluctuate around $0^{\circ}$ and they never exceed $\pm 12^{\circ}$. Whenever we include Faraday rotation, the dispersion varies between $\sim 53^{\circ}$ at low frequencies and $17^{\circ}-35^{\circ}$ at high frequencies. The dispersion at low frequencies is significantly smaller, $< 30^{\circ}$ for the control case (i.e. $\RM = 0$). The edge-on and the face-on view cases including RM converge to the control case at high frequencies. On the other hand, there is no convergence to the control case in the edge-on view even at the highest frequencies that we tested.  We conclude that radio observations can determine the intrinsic orientation of polarisation if the amount of Faraday rotation along the LoS is small. 
  Future observations and RM Synthesis to control the foreground RM, should make it possible to measure the local magnetic field. But we want to stress that this only holds provided the RM along the LoS is sufficiently small. Furthermore, the dispersion of the intrinsic difference, i.e. if $\RM = 0$, is lower if the relic extent along the LoS is  smaller.
 \section{Discussion and conclusion}\label{sec::conclusion}
 In this paper, we studied the polarisation of radio relics in cosmological simulations. This pilot study has the objective to explore the degree of realism of magnetic fields produced in these simulations and to highlight their current limitations. We combined the formalism in \citet{1966MNRAS.133...67B} and \citet{2007MNRAS.375...77H} to model the polarised radio emission of a radio relic. Our simulation grid gave us the possibilities to study the relic seen in six different projections. For each projection, we included {\it internal} and {\it external} Faraday rotation as well as the control case without Faraday rotation. Furthermore, we computed the polarised emission for observing frequencies between $150 \ \MHz$ and $ 4.85 \ \GHz$ and we produced mock observations for the radio telescopes  Effelsberg, VLA, and LOFAR. Our work aims at improving the present knowledge about the following key questions: 
\begin{itemize}
  \item {\it What are the polarisation properties of radio relics?}\\
 The average degree of polarisation of the relic without Faraday Rotation for frequencies above $1.0 \ \GHz$ is  $\ge 50 \ \%$ and, it slightly increases with frequency. These values are close to the intrinsic degree of polarisation. Hence, the depolarisation due to different magnetic field orientations along the LoS inside the emitting region is low. For frequencies below $1.0 \ \GHz$, the average degree of polarisation decreases due to the growing size of the emitting region. \\  
  When including Faraday rotation, the average degree of polarisation drops significantly for low observing frequencies. In this case, the strength of the depolarisation depends on the position of the relic along the LoS and also the Faraday depth of the emitting structure. Whereas for high frequencies, $\ge 1.0-2.0 \ \GHz$, the degree of polarisation is closer to the cases without Faraday rotation. Therefore, we found a polarisation degree significantly larger than the one measured in observations. If the effect of beam depolarisation is included, the degree of polarisation decreases even at high frequencies (i.e. from $\sim 50 \ \%$ to $\sim 25 \ \%$ if the beam resolution goes from $5 \ \arcsec$ to $25 \ \arcsec$  at $1.4 \ \GHz$).\\
  \item {\it What is the RM distribution in simulations?}\\
  We found asymmetric RM distributions with average values of a few $10-100 \ \rad / \m^2$ and standard deviations, $\sigma_{\RM}$, that vary between a few tens $\rad / \m^2$ up to several hundreds $\rad / \m^2$. These values agree with real RM measurements in galaxy clusters. Finally, we found that the shapes of the different RM distributions across our simulated relic are neither Gaussian nor symmetric, which is an usual assumption in several theoretical works \citep[e.g.][]{1966MNRAS.133...67B}. \\
  \item {\it Can simulations reproduce the morphology of the observed polarised emission and does the morphology reflect the magnetic field structure in the relic region?} \\
  The simulated relic does not show the large-scale ($\geq 1-2 \ \Mpc$) polarisation structures observed in real systems such as CIZA J2242.8+5301 or Abell 2744, where the $E$-vectors of the polarisation tightly align with the shock normal. On the other hand, the simulation produces the observed structures on smaller scales ($\leq 200 \ \kpc$), i.e. the strong alignment of $E$-vectors as in CIZA J2242.8+5301, the depolarised "brush" as in 1RXS J0603.3+4214 or a bridge. The latter is produced by a filamentary shock structure that would be visible if observed with sufficiently high resolution. These filaments are similar to the ones observed in 1RXS J0603.3+4214 and CIZA J2242.8+5301. Moreover, we found that the orientation of the $E$-vectors depends on the behaviour of the upstream magnetic field. All these similarities occur on scales that are comparable to the correlation length of the magnetic field \citep[][]{2018MNRAS.474.1672V,2019arXiv190311052D}. \\
 We report that the polarisation $B$-vectors are reasonably aligned with the radio-emission-weighted direction of the magnetic field for $\nu \ge 3.0 \ \GHz$ if one corrects for the effect of Faraday rotation. Hence, despite a few cases in which a large column density along the LoS de-correlates the two vector fields, one can infer the magnetic field structure in the downstream relic region at these frequencies. We found that the magnetic field is neither only parallel nor entirely perpendicular to the shock normal, and therefore the polarised emission reflects the magnetic field and its correlation length limited to regions of a few hundred $\kpc$. \\
  Regardless of the LoS, we found very little to no correlation between magnetic field direction and polarisation $B$-vectors for $\nu < 3.0 \ \GHz$. Additionally, there is an intrinsic misalignment, whose magnitude depends on the extent of the relic and the amount of residual emission along the LoS. \\
  \item  {\it Can simulations reproduce the observed degree of polarisation?} \\
  For frequencies above $>1 \ \GHz$, the average degree of polarisation across the simulated relic is more than $> 50 \ \%$ and, thus, it is significantly larger than in observations. However, the simulated relic shows local peaks of $\sim 70 \ \%$, which is in line with local measurements in observation. Our findings agree with the results of \citet{2013ApJ...765...21S} (the only other numerical simulation using MHD in cosmology we can compare to) who also measured a maximum degree of polarisation of $\approx 75 \ \%$. \\
  A few radio-bright cells in our simulation determine the degree of polarisation and hence, it mirrors the intrinsic properties of these cells. Increasing the numerical resolution would produce a magnetic field that is more tangled on smaller scales. As a consequence, the average degree of polarisation could decrease even if the individual cells have a high degree of polarisation. \\
  The average degree of polarisation is reduced to $4 \ \% - 44 \ \%$ in the mock observations of our relic, depending on redshift, beam size and observing frequency. This is in reasonable agreement with current radio observations. Though, the Effelsberg mock observations do not reproduce the high degree of polarisation measured by \citet{2017A&A...600A..18K} for the relics in CIZA J2242.8+5301 and 1RXS J0603.3+4214. The fact that our results show discrepancies with these observations, is due to either a correlation length in our simulation that is too small or due to missing micro-physics at the shock front. \\  
  \item {\it  Which simulated relic properties are most different from observations?} \\
 The average Mach number of $M>2$ shocks that produces the relic is $\langle M \rangle \sim 2.4$ 
 While the Mach number derived using the spectral index is $\sim 3.5$. Therefore, they are of the same order as estimates from observations \citep[e.g.][]{2018MNRAS.478.2218H}. Our relic is $\sim 1.7 \ \Mpc$ away from the cluster centre, which is similar to the observed relics (see Tab. \ref{tab::relics_obs}). Yet, we noticed that when seen edge-on, its largest-linear size is smaller than most of the observed relics. This projection's extent along the LoS is larger than the estimations for the relic in CIZA J2242.8+5301 \citep[i.e. $250 \ \kpc$ from][]{2017A&A...600A..18K}. Therefore, it is difficult to generalise our findings to the different geometries found in real systems. We require more simulated radio relics in order to better understand the observations. \\
 At the relic position, we found an average magnetic field of $\sim 0.2-0.4 \ \mu \G$ and a radio-weighted magnetic field of $\sim 0.5 \ \mu \G$. These values are consistent with the lower limits estimated in most of the observations \citep[e.g. see Fig. 14 in][]{2013MNRAS.433.3208B}. It is just in the case of Abell 3667 \citet[][]{2010ApJ...715.1143F} where we find that the magnetic field strength obtained in the simulation is in discrepancy with  observations as they report a strength of $\sim 3 \ \mu \G$ in the relic region. \\
 The RM and its dispersion, $\sigma_{\RM}$, are of the same order as of observed RM \citep[e.g.][]{2013MNRAS.433.3208B,2016A&A...596A..22B}. Our work supports that these values vary with respect to the cluster centre depending on two factors: the impact parameter and the specific LoS. In the most extreme cases, the variation can be of a factor of $\sim 10$. 
 Finally, we can say that our simulation produces realistic magnetic field strengths, but in order to reproduce the same $\sigma_{\RM}$ with a higher magnetic field strength (as in the case of Abell 3667), smaller correlation lengths are required.  
\end{itemize}
 Our results show that the polarised emission of radio relics should strongly depend on the environment and the orientation of the polarisation changes with the properties of the upstream gas. The laminar gas flows in the upstream produce a parallel alignment of the $E$-vectors, while disturbances in the upstream will cause a random orientation. This might reflect the local correlation length of the magnetic field. We also found that high-resolution observations above $2.0-3.0 \ \GHz$ will be able to reasonably estimate the magnetic field direction in the relic regions, provided that one corrects for Faraday rotation. In general, this can be done using high-resolution Faraday spectra.\\
 The fluctuation of the magnetic field on small scales points towards a small scale field that is aligned by the shock on microscopic scales to explain the observed degrees of polarisation. We are aware that adding micro-physics to our simulations will help us to understand better the observed degree of polarisation in relics. Nevertheless, this will remain to be a task for a future work. \\
 It is not possible to make any conclusive assessment on the large-scale alignment of polarised emission observed in many radio relics (e.g. CIZA J2242.8+5301, 1RXS J0603.3+4214 etc) due to our small statistical sample. Either they are produced by ordered large-scale magnetic fields, or they are the result of the compressed tangled magnetic fields. The two options are found on scales of $\leq 200 \ \kpc$ in different regions of the same simulated relic. Therefore, more simulated relics (also including strong shocks) are required to generate meaningful statistics. \\
%
 In conclusion, some of our results may quantitatively change if we use a higher spatial resolution, which is difficult to reach at the moment. In particular, some of the small scale details in our relic have the same size as the effective resolution of the Dedner-cleaning applied in our MHD-scheme and, therefore, magnetic structures on scales $\leq 32 \ \kpc$ may be affected by numerical diffusion. Hence, they would show less structures in reality, provided that the magnetic Prandtl number is $P_{\mathrm{m}} \leq 1$ and the magnetic Reynolds number is $R_{\mathrm{m}} \gg 10^2$ in the real ICM. Lastly, the numerical resolution might also affect the correlation length of the magnetic field \citep[e.g.][]{review_dynamo}. \\
 Our study used a model of relativistic electrons that is still relatively crude and that can be improved. First, our model assumes a constant magnetic field and downstream velocity for computing the ageing of the radio emitting electrons in the downstream region. And second, our simulation does not include the injection of relativistic particles and magnetic fields from active galactic nuclei and radio galaxies. In principle, this might affects the shape and size of radio relics as argued by \citet[][]{2017MNRAS.470..240N}. Further uncertainties are related to the assumed acceleration efficiencies and the magnetic field. In this work, we assumed acceleration efficiencies that solely depend on the Mach number, while recent studies \citep[e.g.][]{2014ApJ...783...91C,2017MNRAS.464.4448W,2018ApJ...856...33K} have shown that they can depend on additional parameters such as the shock obliquity. And finally, it is unclear if different magnetic field seeding mechanisms change the magnetic field morphology in the relic's environment. The goal of future studies would be to improve this modelling by including tailored AMR schemes dedicated to increase the resolution at the relic and also to increase the statistics of simulated relics.
  \section*{acknowledgments}
 We thank our anonymous referee for the helpful comments, which helped us improving the quality of our work.
  The cosmological simulations described in this work were performed using the {\enzo} code (http://enzo-project.org), which is the product of a collaborative effort of scientists at many universities and national laboratories. We gratefully acknowledge the {\enzo} development group for providing extremely helpful and well-maintained on-line documentation and tutorials. \\
  The authors gratefully acknowledge the Gauss Centre for Supercomputing e.V. (www.gauss-centre.eu) for supporting this project by providing computing time through the John von Neumann Institute for Computing (NIC) on the GCS Supercomputer JUWELS at J\"ulich Supercomputing Centre (JSC), under projects no. 11823, 10755 and 9016 as well as hhh42,  hhh44 and stressicm. \\ 
The original simulations on which this work is based have been produced under project HHH42 at JSC by F.V. as PI. 
  D. W.,  F.V. and P. D. F. acknowledge financial support from the European Union's Horizon 2020 program under the ERC Starting Grant "MAGCOW", no. 714196. We also acknowledge the usage of online storage tools kindly provided by the Inaf Astronomica Archive (IA2) initiave (http://www.ia2.inaf.it).   \\
  MH acknowledges support by the BMBF Verbundforschung under the grant 05A17STA. \\
  We acknowledge fruitful scientific discussion with A. Bonafede, K. Rajpurohit, C. Stuardi, G. Brunetti and S. O'Sullivan.
  This research made use of the radio astronomical database galaxyclusters.com, maintained by the Observatory of Hamburg. 

 \bibliographystyle{mnras}
 \bibliography{mybib}
 
 \appendix
 
  \section{Synchrotron Emission}\label{app::synchrotron}
 In this section, we summarise the mathematical details used to compute the downstream profile of the radio emission (see Sec. \ref{ssec::synchrotron}). Following the approach of \citet{2007MNRAS.375...77H}, we compute the emission per volume and its parallel and perpendicular components at a distance $x$ away from the shock front as the convolution of the electron spectrum $n_{\mathrm{E}}(\tau, x)$ and the function $F(1/\tau^2)$: 
 \begin{align}
     \frac{\dd P}{\dd V \dd \nu }(x) &= C_{\mathrm{R}}  \int_{0}^{E_{\max}} n_{\mathrm{E}}(\tau, x) F\left(\frac{1}{\tau^2}\right) \dd \tau  \label{eq::dPdVdv1} \\
     \frac{\dd \Ppara}{\dd V \dd \nu }(x) &= C_{\mathrm{R}} \int_{0}^{E_{\max}} n_{\mathrm{E}}(\tau, x) \left[F\left(\frac{1}{\tau^2}\right)-G\left(\frac{1}{\tau^2}\right)\right]\dd \tau \label{eq::ppara1} \\
     \frac{\dd \Pperp}{\dd V \dd \nu }(x) &= C_{\mathrm{R}} \int_{0}^{E_{\max}} n_{\mathrm{E}}(\tau, x) \left[F\left(\frac{1}{\tau^2}\right)+G\left(\frac{1}{\tau^2}\right)\right]\dd \tau \label{eq::pperp1} .
 \end{align}
 The functions $F(x)$ and $G(x)$ depend on the modified Bessel functions $K$ as \citep[see][]{rybickiandlightman}:
 \begin{align}
  F(x) &= x \int_x^{\infty} K_{\frac{5}{3}}(\xi) \dd \xi  \\
  G(x) &= x K_{\frac{2}{3}}(x) .
 \end{align}    
 The constant $C_{\mathrm{R}}$ is computed as:
 \begin{align}
     C_{\mathrm{R}} &= \frac{9 e^{5/2} B^{3/2} \sin{\alpha}}{4\sqrt{\nuobs \me c}},
 \end{align}
 where $\me$ and $e$ the electron mass and charge, $B$ is the magnetic field, $\alpha$ is the pitch angle, $\nuobs$ is the observing frequency, $c$ is the speed of light and $E$ is the electron energy. The function $\tau$ depends on the energy as:
 \begin{align}
     \tau &= \sqrt{\frac{3 e B}{16 \nuobs \me c}} \left(\frac{E}{\me c^2}+1\right).
 \end{align}
 The electron spectrum at a distance $x$ to the shock is computed as:
 \begin{align} 
   \begin{split}
      n_{\mathrm{E}}&(E, x) =  \frac{n_{\mathrm{e}} C_{\mathrm{spec}} }{\me c^2} \left(\frac{E}{\me c^2}\right)^{-s}   \\
      &\times \left[1-\left(\frac{\me c^2}{E_{\max}} + C_{\mathrm{cool}}\frac{x}{v_d}\right)\frac{E}{\me c^2} \right]^{s-2}.
     \end{split} \label{eq::ne1}
 \end{align}
 Electrons can only be accelerated to a finite energy $E_{\max}$. Hence, the spectrum is evaluated if $EC_{\mathrm{cool}} x /v_d / \me / c^2< 1 - E/E_{\max}$, with $v_d$ being the downstream velocity of the shock. The cooling constant is given as:
 \begin{align}
     C_{\mathrm{cool}} = \frac{\sigma_{\mathrm{T}}}{6 \me c\pi} \left(B^2_{\mathrm{CMB}}+B^2\right).
 \end{align}
 Here, $\sigma_{\mathrm{T}}$ is the Thomson cross-section and $B^2_{\mathrm{CMB}}$ is the equivalent magnetic field of the cosmic microwave background at redshift $z$. The normalisation of the spectrum is:
 \begin{align}
     \Cspec = \xi_{\mathrm{e}} \frac{u_d}{c^2}\frac{\mpr}{\me}\frac{q-1}{q}\frac{1}{I_{\mathrm{spec}}},
 \end{align}
 where $\xi_{\mathrm{e}}$ is the acceleration efficiency, $q$ is the entropy jump across the shock, $u_d$ is the internal energy of the downstream gas and $\mpr$ is the proton mass. The integral $I_{\mathrm{spec}}$ is given as
 \begin{align}
     I_{\mathrm{spec}} = \int_{E_{\min}}^{\infty} E \left(\frac{E}{\me c^2}\right)^{-s} \left(1-\frac{E}{E_{\max}}\right)^{s-2} \dd E.
 \end{align}
 For the minimum energy, above which electrons are considered to be suprathermal, we chose $E_{\min} = 10 k_{\mathrm{b}} T$.
 \section{Average of angles}\label{app::angles}
 We use Eq. \ref{eq::radio_weighting} to compute the radio-weighted average of a physical quantity $Q$. If $Q$ is an angle $\alpha$, we first compute the two components: $x = \sin ( \alpha) $ and $y = \cos (\alpha)$. Next, we compute the radio-weighted averages of the two components $\langle x \rangle_{\nuobs}$ and $\langle y \rangle_{\nuobs}$. Finally, we compute the radio-weighted average angle as:
 \begin{align}
     \langle \alpha \rangle_{\nuobs} = \arctan \left( \frac{\langle x \rangle_{\nuobs}}{\langle y \rangle_{\nuobs}} \right).
 \end{align}
This approach is reasonable, if the scatter of angles is about $90^{\circ}$ or less. In our case, this works well for almost all LoS, since a few cells dominate the radio emission. We applied this algorithm, when we computed the radio-weighted orientation of the physical magnetic field (see Sec. \ref{ssec::magnetic}). This is not to be confused with the orientation of the polarisation vectors which we computed using Eq. \ref{eq::burn} and Eq. \ref{eq::beta}.
  \section{The correlation of the magnetic spectrum}
\label{app:1}

The magnetic power spectrum, in a magnetised plasma, has a characteristic shape that differs from the shape of the kinetic power spectrum.  In detail, it cannot be characterised only by a power-law. The galaxy cluster produced by our MHD simulation has enough resolution for better resolving the morphology of magnetic fields at small-scales during structure formation. Therefore, we can better constrain the shape of the magnetic power spectrum. The magnetic spectral properties of the cluster has already been analysed in \citet{2019arXiv190311052D}. One important result of this work is that despite of the different dynamical states, the magnetic spectra can be well-fitted by the following equation:
\begin{align}\label{fit_eq}
E_M(k)= A \,k^{3/2}\left[ 1- \text{erf} \left[ B \ln\left( \frac{k}{C} \right)  \right]  \right]
\end{align}
where $A$ is the normalisation, $B$ is related to the width of the spectra and $C$  is a characteristic wavenumber which inverse corresponds to the inverse outer scale of the magnetic field. In this work, we computed the power spectrum of our data and then fitted it to Eq. \ref{fit_eq}. In this work, we refer to the correlation length given by the outer scale of the magnetic spectrum, i.e. the $C$ parameter.

\end{document}